\documentclass[a4paper,11pt]{article}
\pdfoutput=1 

\usepackage{jheppub} 

\usepackage[T1]{fontenc} % if needed

\usepackage{amsmath,amssymb,amsfonts,amsthm,dsfont,dcolumn,yfonts,bm,simplewick}
\usepackage{graphicx,wrapfig}

\usepackage{mathrsfs}
\usepackage{empheq}

\edef\restoreparindent{\parindent=\the\parindent\relax}
\usepackage{parskip}
\restoreparindent

\allowdisplaybreaks[1]

\newcommand{\cN}{\mathcal{N}}

\newcommand{\cJ}{\mathcal{J}}
\newcommand{\cK}{\mathcal{K}}
\newcommand{\sk}{\kappa} %shifted level k
\newcommand{\sltwof}{\mathfrak{sl}(2,\mathds{R})^{(1)}}
\newcommand{\sltwo}{\mathfrak{sl}(2,\mathds{R})}
\newcommand{\sutwof}{\mathfrak{su}(2)^{(1)}}
\newcommand{\sutwo}{\mathfrak{su}(2)}

\numberwithin{equation}{section}

\title{\boldmath Higher spins on AdS$_{3}$ from the worldsheet}

\author{Kevin Ferreira, Matthias R.\ Gaberdiel and Juan I.\ Jottar}
\affiliation{Institut f\"ur Theoretische Physik, ETH Zurich,\\
 CH-8093 Z\"urich, Switzerland}

\emailAdd{kevinp@phys.ethz.ch}
\emailAdd{gaberdiel@itp.phys.ethz.ch}
\emailAdd{jijottar@phys.ethz.ch}

\abstract{It was recently shown that  the CFT dual of string theory on ${\rm AdS}_3 \times {\rm S}^3 \times \mathds{T}^4$, 
the symmetric orbifold of $\mathds{T}^4$, contains a closed higher spin subsector. Via holography, this
makes precise the sense in which tensionless string theory on this background contains a Vasiliev higher spin theory. 
In this paper we study this phenomenon directly from the worldsheet. Using the WZW description of the 
background with pure NS-NS flux, we identify the states that make up the leading Regge trajectory and show that
they fit into the even spin ${\cal N}=4$ Vasiliev higher spin theory. We also show that these higher spin states
do not become massless, except for the somewhat singular case of level $k=1$ where the theory
contains a stringy tower of massless higher spin fields coming from the long string sector.}

\begin{document} 
\maketitle
\flushbottom

%%%%%%%%%%%%%%%%%%%%%%%%%%%%%%%%%%%%%%%%%%%%%%%%%%%%%%%%%%%%%%%%%%%%
\section{Introduction}\label{sec:intro}
%%%%%%%%%%%%%%%%%%%%%%%%%%%%%%%%%%%%%%%%%%%%%%%%%%%%%%%%%%%%%%%%%%%%

In the tensionless limit string theory is expected to exhibit a large underlying symmetry that is believed to 
lie at the heart of many special properties of stringy physics \cite{Gross:1988ue,Witten:1988zd,Moore:1993qe}.
In flat space, the tensionless limit is somewhat subtle since there is no natural length scale relative to which the (dimensionful)
string tension may be taken to zero. The situation is much better in the context of string
theory on an AdS background, since the cosmological constant of the AdS space defines a natural length scale. This
is also reflected by the fact that higher spin theories -- they are believed to capture the symmetries of the
leading Regge trajectory at the tensionless point \cite{Sundborg:2000wp,Mikhailov:2002bp} --- 
appear naturally in AdS backgrounds \cite{Vasiliev:2003ev}. 

In the context of string theory on AdS$_3$ concrete evidence for this picture was recently
obtained in \cite{Gaberdiel:2014cha}. More specifically, it was shown that the CFT dual of 
string theory on ${\rm AdS}_3\times {\rm S}^3 \times \mathds{T}^4$, the symmetric
orbifold of $\mathds{T}^4$, see \cite{David:2002wn} for a review, contains the CFT
dual of the supersymmetric higher spin theories constructed in \cite{Gaberdiel:2013vva}.\footnote{Superconformal 
higher spin theories were first constructed in \cite{Prokushkin:1998bq,Prokushkin:1998vn}.
The duality is the natural supersymmetric analogue of the original bosonic 
proposal of \cite{Gaberdiel:2010pz}, see \cite{Gaberdiel:2012uj} for a review; various properties of the ${\cal N}=4$ duality 
were further analysed in \cite{Beccaria:2014jra,Gaberdiel:2014yla}.}
While this indirect evidence is very convincing, it would be very interesting to have
more direct access to the higher spin sub-symmetry in string theory. This symmetry
is only expected to emerge in the tensionless limit of string theory, in which the string
is very floppy and usual supergravity methods are not reliable. Thus we should attempt to 
address this question using a worldsheet approach.

Worldsheet descriptions of string theory on AdS backgrounds are notoriously hard, but 
in the context of string theory on AdS$_3$, the background with pure NS-NS flux admits
a relatively straightforward worldsheet description in terms of a WZW model based on 
the Lie algebra $\mathfrak{sl}(2,\mathds{R})$ \cite{Maldacena:2000hw,Maldacena:2000kv,Maldacena:2001km}. 
In this paper we shall use this approach to look for signs of a higher spin symmetry among
these worldsheet theories. More concretely, we shall combine the WZW model corresponding to 
$\mathfrak{sl}(2,\mathds{R})$  with an $\mathfrak{su}(2)$ WZW model, describing strings propagating
on ${\rm S}^3$, as well as four free fermions and bosons corresponding to $\mathds{T}^4$. The complete
critical worldsheet theory then describes strings on ${\rm AdS}_3\times {\rm S}^3 \times \mathds{T}^4$. 

The worldsheet description of these WZW models contains one free parameter, the level $k$ 
of the ${\cal N}=1$ superconformal WZW models associated to $\mathfrak{sl}(2,\mathds{R})$ and 
$\mathfrak{su}(2)$, respectively --- these two levels have to be the same
in order for the full theory to be critical. Geometrically, these levels correspond to the 
size of the AdS$_3$ space (and the radius of ${\rm S}^3$) in string units. The tensionless limit should therefore
correspond to the limit where $k$ is taken to be small. 

In this paper we analyse systematically the string spectrum of the worldsheet theory for $k$ small.\footnote{See e.g.\ 
\cite{Isberg:1993av,Sagnotti:2003qa,Bagchi:2016yyf,Bagchi:2015nca} and references therein for other worldsheet approaches 
to this problem which do not focus on the spectrum itself, but rather on the symmetry structures that are presumed to 
emerge in the tensionless limit of string theory.} As 
we shall show, the only massless spin fields that emerge in this limit are those associated to the 
supergravity multiplet, while all the higher spin fields remain massive, except in the extremal case where the level is 
taken to be $k=1$ --- this is strictly speaking an unphysical value for the level since then the bosonic $\mathfrak{su}(2)$
model has negative level; however, as argued in \cite{Seiberg:1999xz}, some aspects of the theory may still make sense. 
(We should also mention that in the context of the WZW model based on ${\rm AdS}_3\times {\rm S}^3 \times {\rm S}^3 \times {\rm S}^1$
\cite{Elitzur:1998mm,Eberhardt:2017fsi}
the theory with $k=1$ is not singular since it is compatible with the levels of the two superconformal
$\mathfrak{su}(2)$ models being $k^+=k^-=2$, leading to vanishing bosonic levels for the two 
$\mathfrak{su}(2)$ algebras.)\footnote{We 
thank Lorenz Eberhardt for a useful discussion about this point.}
For $k=1$, the bosonic $\mathfrak{sl}(2,\mathds{R})$ algebra has level $k_{\rm bos}=3$,
and as in \cite{GGH}, an infinite tower of massless higher spin fields arises from the long string 
subsector (the spectrally flowed continuous representations). These higher spin fields are part of a continuum 
and realise quite explicitly some of the speculations of \cite{Seiberg:1999xz}. 

\begin{center}
\begin{figure}[h!]
%\hspace*{-1.0cm} \includegraphics[scale=0.34]{Regges.pdf}
 \hspace*{-2cm}\includegraphics[width=19cm]{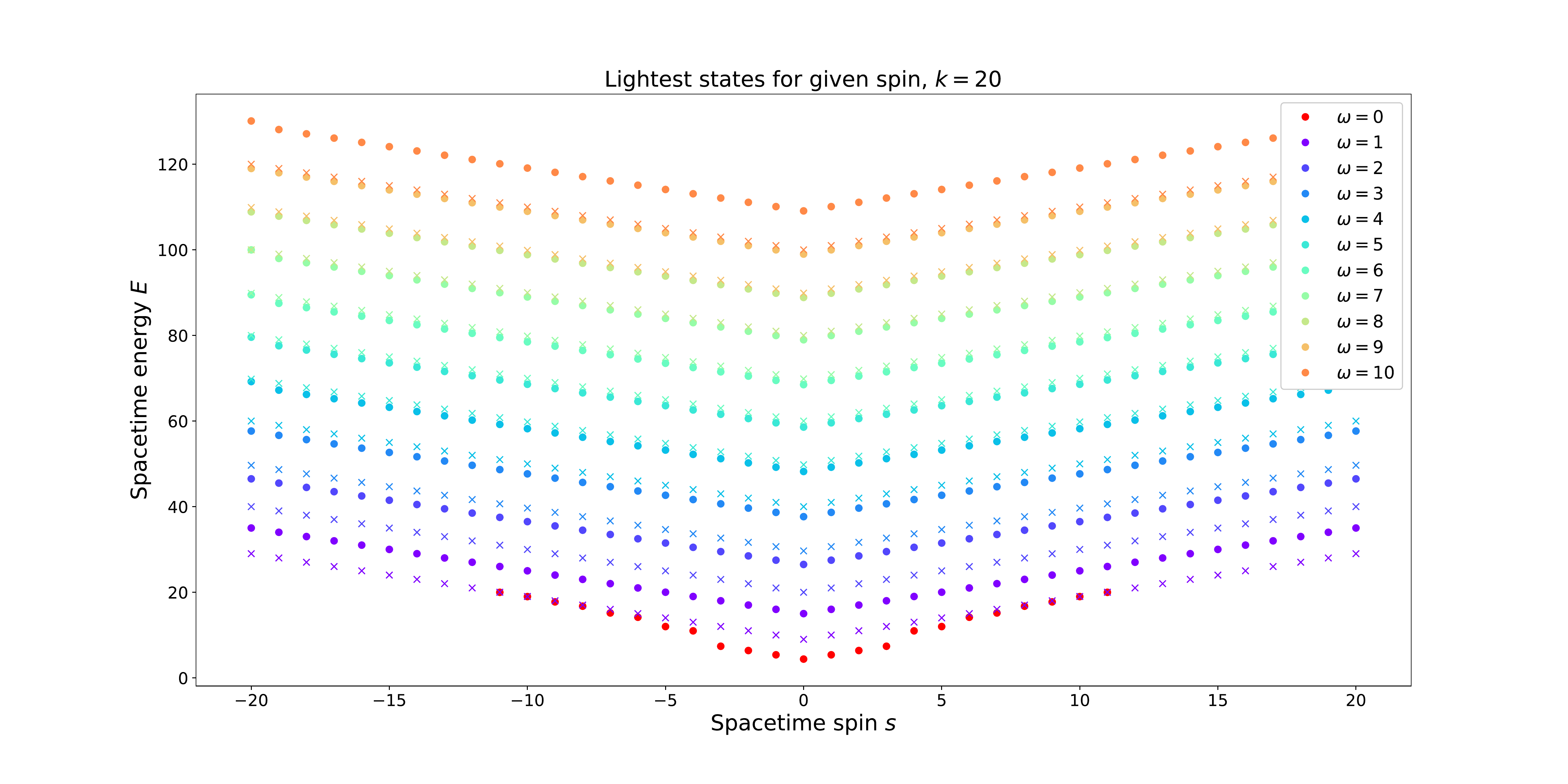}
\caption{Lowest energy states for fixed $k=20$ as a function of the spin $s$. Crosses denote continuous representations (long strings), 
whereas dots correspond to discrete representations. The red dots describe the unflowed discrete states, while the 
different bands correspond, in turn, to the flows $\omega=1,2,\ldots,10$. The lines corresponding to the spectrally flowed representations have 
been artificially capped for illustration purposes; they continue all the way to $s=\pm\infty$.\label{Regges}}
\end{figure}
\end{center}

For more generic values of the level, we also explain the sense in which 
a `leading Regge trajectory'
emerges, and we give an explicit description of these states. In particular, we show that
the relevant states form the spectrum of  a specific ${\cal N}=4$ higher spin theory of Vasiliev
that was recently analysed in detail by one of us \cite{Ferreira:2017zbh}. (More specifically, this higher
spin theory consists of one ${\cal N}=4$ multiplet for each even spin; the fact that the leading Regge 
trajectory in closed string theory only consists of states (or multiplets) of even spin is also familiar from flat space, 
see the discussion around eq.~(\ref{flatspace}).) 
For spins that are small
relative to the size of the AdS space, the states on the leading Regge trajectory are described by physical
states coming from the (unflowed) discrete representations of $\mathfrak{sl}(2,\mathds{R})$; as the spin gets larger,
the corresponding classical strings become longer until they hit the boundary of the AdS space where they
become part of the spectrally flowed continuous representations, describing the continuum of long strings, see
Figure~\ref{Regges}. 
\noindent This picture fits in nicely with expectations from \cite{Maldacena:2000hw,Maldacena:2000kv,Maldacena:2001km}, see
also \cite{Seiberg:1999xz}. 

The fact that among these backgrounds with NS-NS flux no conventional higher spin symmetry emerges also has
a natural interpretation in terms of the structure of the classical sigma model. Indeed, as explained in 
\cite{Berkovits:1999im}, the tension of the string is of the form \cite[eq.~(7.34)]{Berkovits:1999im} 
\begin{equation}\label{Tensioneq}
T = \sqrt{ Q_{\text{NS}}^2 + g_s^2 \, Q_{\text{RR}}^2} \ , 
\end{equation}
where $Q_{\text{NS}}$ and  $Q_{\text{RR}}$ are quantized, and $g_s$ is the string coupling constant. This formula therefore suggests that
the tensionless limit is only accessible in the situation with pure R-R flux (and in the limit $g_s\rightarrow 0$).
\medskip

The paper is organized as follows. We explain the basics of the worldsheet theory (and set up our
notation) in Section~2. In Section~3 we prove that the spectrum of this family of worldsheet theories 
does not contain any massless
higher spin fields among the unflowed representations (describing short strings). 
In Section~4 we start with identifying the states that comprise the leading Regge 
trajectory. We first analyse the states of low spin that arise from the unflowed discrete representations. We 
also comment on the structure of the subleading Regge trajectory, as well as the situation for the case where 
$\mathds{T}^4$ is replaced by K3. The rest of the leading Regge trajectory that is part of the 
continuous spectrum is then identified in Section~5. We also comment there on the massless higher spin fields
arising from the spectrally flowed continuous representation at $k=1$, and explain how they fit in with the 
expectations from \cite{Seiberg:1999xz}. 
Section~6 contains our conclusions, and there are three
appendices where we have collected some of the more technical arguments that are referred to at various places in 
the body of the paper.

%%%%%%%%%%%%%%%%%%%%%%%%%%%%%%%%%%%%%%%%%%%%%%%%%%%%%%%%%%%%%%%%%%%%
\section{Worldsheet string theory on AdS$_{3}$}
%%%%%%%%%%%%%%%%%%%%%%%%%%%%%%%%%%%%%%%%%%%%%%%%%%%%%%%%%%%%%%%%%%%%

We want to study the spectrum of type IIB strings propagating on backgrounds of the form AdS$_{3}\times {\rm S}^{3} \times X$, where
$X$ is either $\mathds{T}^{4}$ or ${\rm K3}$ so that the resulting theory has $\cN=4$ spacetime supersymmetry. We shall concentrate
on the background with pure NS-NS flux for which  the AdS$_{3}\times {\rm S}^{3}$ theory can be described by a (non-compact) 
${\rm SL}(2,\mathds{R})\times {\rm SU}(2)$ WZW model\footnote{Strictly speaking, we always consider the universal cover of the 
${\rm SL}(2,\mathds{R})$ group, because the timelike direction of AdS$_{3}$ is taken to be non-compact.} that can be studied by 
conventional CFT methods. The bosonic version 
of this theory was discussed in some detail in the seminal papers \cite{Maldacena:2000hw,Maldacena:2000kv,Maldacena:2001km}; 
in what follows we extend, following \cite{Giveon:1998ns,Pakman:2003cu,Israel:2003ry,Raju:2007uj},  some aspects of their analysis to the 
supersymmetric case. 

The symmetry algebras of the supersymmetric WZW models are the $\cN=1$ superconformal affine algebras 
$\sltwof_{k} \oplus \sutwof_{k'}$ that will be described in more detail below. Their central charges equal
\begin{equation}\label{ccharges}
c ( \sltwof_{k} ) = 3 \Bigl( \frac{k+2}{k} + \frac{1}{2} \Bigr) \ , \qquad
c ( \sutwof_{k'} ) = 3 \Bigl( \frac{k'-2}{k'} + \frac{1}{2} \Bigr) \ ,
\end{equation}
and the condition that the total charge adds up to $c=9$ (as befits a $6$-dimensional supersymmetric background) 
requires then that $k=k'$. For this choice of levels, the naive $\cN=1$ worldsheet supersymmetry of the model is 
enhanced to $\cN=2$ \cite{Ivanov:1994ec,Giveon:1998ns}. This enhancement can also be understood from the fact that the 
AdS$_{3}\times {\rm S}^{3}$ theory can be described as a non-linear sigma model on the supergroup 
PSL$(2\vert 2)$ (see, e.g., \cite{Gerigk:2012lqa}).

%%%%%%%%%%%%%%%%%%%%%%%%%%%%%%%%%%%
\subsection{The AdS$_3$ WZW model}\label{sec:AdS3WZW}
%%%%%%%%%%%%%%%%%%%%%%%%%%%%%%%%%%%
In our conventions, the $\sltwof$ algebra describing superstrings on AdS$_{3}$ reads
\begin{alignat}{5}
\bigl[J^{+}_{m},J^{-}_{n}\bigr] 
={}&
 -2J^{3}_{m+n} + km\delta_{m,-n}&
 & \quad &
 \bigl[J^{3}_{m},J^{\pm}_{n}\bigr] 
={}&
 \pm J^{\pm}_{m+n}&
  & \quad &
 \bigl[J^{3}_{m},J^{3}_{n}\bigr] 
={}&
 -\frac{k}{2}m\delta_{m,-n}
 \nonumber\\
 \bigl[J^{\pm}_{m},\psi^{3}_{r}\bigr]
={}&
 \mp\psi^{\pm}_{m+r} &
 & \quad &
 \bigl[J^{3}_{m},\psi^{\pm}_{r}\bigr]
={}&
 \pm\psi^{\pm}_{m+r}&
  &\quad &
 \bigl[J^{\pm}_{m},\psi^{\mp}_{r}\bigr]
 ={}&
 \mp 2\psi^{3}_{m+r}
 \\
 \bigl\{\psi^{+}_{r},\psi^{-}_{s}\bigr\}
={}&
 k\delta_{r,-s}&
 & \quad &
 \bigl\{\psi^{3}_{r},\psi^{3}_{s}\bigr\}
={}&
 -\frac{k}{2}\delta_{r,-s}\,.  &
 \nonumber
\end{alignat}
The dual Coxeter number is $h^\vee_{\sltwo} = -2\,$. As detailed in appendix \ref{app: SUSY Kac Moody}, the shifted currents 
\begin{align}
\cJ^{+} 
={}&
J^{+} + \frac{2}{k}\bigl(\psi^{3}\psi^{+}\bigr) 
\nonumber\\
\cJ^{-} 
={}&
J^{-} -\frac{2}{k}\bigl(\psi^{3}\psi^{-}\bigr) \label{Jcal}
\\
\cJ^{3} 
={}&
J^{3} +\frac{1}{k}\bigl(\psi^{-}\psi^{+}\bigr) 
\nonumber
\end{align}
decouple from the fermions, $\left[\cJ^{a}_{n},\psi^{b}_{r}\right]=0\,$, and satisfy the same algebra as the $J^{a}$ with level $\kappa = k+2\,$. The Sugawara stress tensor and supercurrent are 
\begin{align}
T
={}&
 \frac{1}{2k}\left(\cJ^{+}\cJ^{-}+\cJ^{-}\cJ^{+} -2\cJ^{3}\cJ^{3}-\psi^{+}\partial \psi^{-}-\psi^{-}\partial\psi^{+}+2\psi^{3}\partial\psi^{3}\right)
\label{Tdef} \\
 G
 ={}&
 \frac{1}{k}\left(\cJ^{+}\psi^{-}+\cJ^{-}\psi^{+} -2\cJ^{3}\psi^{3} - \frac{2}{k}\psi^{+}\psi^{-}\psi^{3}\right)\ ,
 \label{Gdef}
\end{align}
where every composite operator in the above expressions is understood to be normal-ordered. These generators satisfy the 
$\cN=1$ superconformal algebra 
\eqref{N equals 1 superconformal algebra 1}--\eqref{N equals 1 superconformal algebra 3} with central charge (see eq.~\eqref{total central charge})
\begin{equation}
c  = 3\left(\frac{k+2}{k}+\frac{1}{2}\right)\ .
\end{equation}
The holographic dictionary implies that the global charges in the spacetime theory are given by \cite{Giveon:1998ns}
\begin{equation}\label{spacetime charges}
L^{\text{CFT}}_{0} = J^{3}_{0}\ ,\qquad L^{\text{CFT}}_{1} = J^{-}_{0}\ ,\qquad L^{\text{CFT}}_{-1} = J^{+}_{0}\ ,
\end{equation}
with analogous expressions for the right-movers. In particular, the spacetime conformal dimension (which we henceforth refer to as the energy $E$) 
is given by the eigenvalue of $J^{3}_{0} + \bar{J}^{3}_{0}$, while the spacetime helicity $s$ equals 
$J^{3}_{0} - \bar{J}^{3}_{0}$.\footnote{In the following, we shall refer to $s$ as the spin --- this is what it is from the 
view of the two-dimensional spacetime conformal field theory.}
Since we want to keep track of these quantum numbers, it will prove convenient to describe the representation content 
with respect to the coupled currents $J^{a}$. 

In addition to the symmetry algebra, the actual worldsheet conformal field theory is characterised by the spectrum, i.e., 
by the set of $\sltwof$ representations that appear in the theory. For the bosonic case, a proposal for what this spectrum
should be was made in  \cite{Maldacena:2000hw}, and the same arguments also apply here once we decouple the fermions. Recall that a highest weight representation of a (bosonic) affine 
Kac-Moody algebra is uniquely characterised by the representation of the zero mode algebra (in our case $\mathfrak{sl}(2,\mathds{R})$) 
acting on the `ground states' --- these are the states that are annihilated by the modes ${\cal J}^a_n$ with $n>0$. For the case at hand, the 
relevant representations of $\mathfrak{sl}(2,\mathds{R})$ that appear \cite{Maldacena:2000hw} are the so-called principal discrete representations 
(corresponding to short strings), as well as the principal continuous representations --- together they form a 
complete basis of square-integrable functions on AdS$_{3}$.  Furthermore, since the no-ghost theorem truncates the 
set of these representations to a finite number (depending on $k$) \cite{Hwang:1990aq,Evans:1998qu}, additional 
representations corresponding to their spectrally flowed images appear  
\cite{Maldacena:2000hw}; these describe the long strings. In each case, the representation on the ground states is the 
same for left- and right-movers --- this
theory is therefore the natural analogue of the `charge-conjugation' modular invariant, see also \cite{Maldacena:2000kv}. 

In the supersymmetric case we are interested in, we consider the above $\mathfrak{sl}(2,\mathds{R})$ affine theory
for the decoupled bosonic currents  ${\cal J}^a$, and tensor to it a usual free fermion theory (where the 
fermions will either all be in the NS or in the R sector). Note that this will lead to a modular invariant spectrum since both factors 
are separately modular invariant.  

In the following we shall study the spacetime spectum of this worldsheet theory with a view towards identifying
the states on the leading Regge trajectory. We shall first concentrate on the unflowed discrete representations, 
where the low-lying states of the leading Regge trajectory --- those whose spin satisfies $s \lessapprox \frac{k}{2}$ --- 
originate from. The remaining states of the Regge trajectory are part of the continuum of long strings that is described
by the (spectrally flowed) continuous representations; they will be analysed in Section~\ref{sec: spectral}.

\subsubsection{The NS sector} 

In the NS sector we label the ground states by $|j,m\rangle$, where $m$ is the eigenvalue of $J_{0}^{3}$,
while $j$ labels the spin, 
\begin{equation}\label{unshifted Casimir}
C_{2}|j,m\rangle = -j(j-1)|j,m\rangle \,,\qquad J_{0}^{3}|j,m\rangle = m|j,m\rangle\ , 
\end{equation}
and $C_2$ is the quadratic Casimir of $\mathfrak{sl}(2,\mathds{R})$ 
\begin{equation}
C_{2} = \frac{1}{2}\left(J_{0}^{+}J_{0}^{-} + J_{0}^{-}J_{0}^{+}\right)-J_{0}^{3}J_{0}^{3}\ .
\end{equation}
The condition to be ground states, i.e., to satisfy 
\begin{equation}\label{Affine hws}
J^{a}_{n}|j,m\rangle = 0 \quad \forall\, \,n \geq 1 \qquad \text{and}\qquad \psi^{a}_{r}|j,m\rangle = 0 \quad \forall \,\, r \geq \tfrac{1}{2}
\end{equation}
implies, in particular, that the coupled and decoupled bosonic modes with $n\geq 0$ agree on the ground
states,
\begin{equation}\label{JJground}
J^a_n \, |j,m\rangle = {\cal J}^a_n\, |j,m\rangle \ , \qquad n\geq 0 \ ; 
\end{equation}
the correction terms involve positive fermionic modes that annihilate the ground states. (Thus it makes no difference
whether we label the ground states in terms of the decoupled or coupled spins). Furthermore, the ground states are 
annihilated by  
\begin{equation}
L_{n}|j,m\rangle =0 \quad \text{for } n \geq 1 \qquad \text{and}\qquad G_{r}|j,m\rangle = 0 \quad \text{for } r\geq \tfrac{1}{2}\ ,
\end{equation}
as follows from eqs.~(\ref{Tdef}) and (\ref{Gdef}).

The discrete lowest weight representations $\mathcal{D}^{+}_{j}$ --- in \cite{Maldacena:2000hw} they are called `positive energy' --- 
are characterised by the conditions 
\begin{equation}\label{sl2 states normalization}
J_{0}^{+}|j,m\rangle = |j,m+1\rangle\,,\qquad J_{0}^{-}|j,m\rangle = \Bigl(-j(j-1)+m(m-1)\Bigr)|j,m-1\rangle\ .
\end{equation}
Note that the state $|j,j\rangle$ has the lowest $J_{0}^{3}$ eigenvalue and is therefore annihilated by $J_{0}^{-}$. In particular, it follows from \eqref{spacetime charges} that 
\begin{equation}
\mathcal{D}^{+}_{j}:\qquad J_{0}^{-}|j,j\rangle = 0 \quad \Rightarrow \quad L_{1}^{\text{CFT}}|j,j\rangle = 0
\end{equation}
as appropriate for a quasiprimary state in the dual $2d$ CFT. The representation of the full affine algebra is obtained by the action
of the negative modes $J^a_{-n}$ and $\psi^a_{-r}$, acting on these ground states. With respect to the global $\sltwo$ algebra,
all of these states will then also sit in discrete lowest weight representations of $\sltwo$, and the 
quasiprimary states of the dual CFT will always correspond to the lowest weight states of these discrete representations.

\subsubsection{The R sector}

The analysis in the Ramond sector is slightly more subtle since there are fermionic zero modes. 
The ground states are therefore characterised in addition by an irreducible spinor representation of 
the Clifford algebra in $(2+1)$-dimensions, spanned by the fermionic zero modes --- this representation 
is two-dimensional and can be described by $\vert s_0\rangle$, with $s_0=\pm 1$. The full set of 
ground states is therefore labelled by $\vert j,m;s_0\rangle$. 
The presence of the fermionic zero modes implies that, unlike (\ref{JJground}), 
the action of the decoupled and coupled bosonic zero modes differs on the ground states. In particular, 
\begin{equation}
\mathcal{J}_0^3 = J_0^3 - \frac{1}{k}\left(\psi^{+}\psi^{-}\right)_{0}\ ,
\end{equation}
where on the ground states (see eq.~\eqref{Rordering})
\begin{equation}
(\psi^+\psi^-)_0 \, \vert j,m;s_0\rangle = \frac{1}{2}\left[ \psi_0^+, \psi_0^- \right] \vert j,m;s_0\rangle = k\frac{\sigma^3}{2}\vert j,m;s_0\rangle 
= k\frac{s_0}{2} \vert j,m;s_0\rangle \ ,
\end{equation}
and consequently
\begin{align}\label{J0shift}
J_{0}^{3}\vert j,m;s_0\rangle
 = \left( m+\frac{s_0}{2}\right) \vert j,m;s_0\rangle \ .
\end{align}
Effectively, this can be interpreted as shifting the spin $j$ (with respect to the coupled algebra) 
of the R sector representation by $\pm \frac{1}{2}$ relative to the decoupled algebra.

We are interested in organising the descendants of these ground states in terms of representations of the (coupled) 
$\mathfrak{sl}(2,\mathds{R})$ zero modes since they have a direct interpretation in terms of the dual CFT, see 
eq.~(\ref{spacetime charges}). Since the creation generators --- the negative bosonic and fermionic modes --- transform
in the adjoint representation of this $\mathfrak{sl}(2,\mathds{R})$, the spins that arise will be of the form
$j+\ell$, where $j$ is the spin of the (decoupled) ground states while $\ell\in \mathds{Z}$ in the NS sector and 
$\ell\in\mathds{Z}+\frac{1}{2}$ in the R-sector --- here we have absorbed the above shift by $\frac{1}{2}$ into the definition of 
$\ell$. A similar consideration applies for the right-movers where the resulting spin will be $j+\bar{\ell}$ for the same $j$
(and with the same restrictions on $\bar\ell$). 
Thus the total energy and spin of
such a descendant will be 
\begin{equation}\label{ellell}
E = 2 j + \ell + \bar\ell \ , \qquad s= \ell - \bar\ell \ .
\end{equation}
Note that in the NS-NS and R-R sectors the spacetime spin $s$ will be integer, while in the NS-R and R-NS sectors
it will be half-integer. 

%%%%%%%%%%%%%%%%%%%%%%%%%%%%%%%%%%%
\subsection{The compact directions}
%%%%%%%%%%%%%%%%%%%%%%%%%%%%%%%%%%%
The remaining spacetime directions are described by ${\rm S}^3 \times \mathds{T}^4$. Supersymmetric 
strings propagating on ${\rm S}^{3}$ can be described by a WZW model based on $\sutwof$, for which our conventions are
\begin{alignat}{5}
\bigl[K^{+}_{m},K^{-}_{n}\bigr] 
={}&
 2K^{3}_{m+n} + km\delta_{m,-n}&
 &\quad &
 \bigl[K^{3}_{m},K^{\pm}_{n}\bigr] 
={}&
 \pm K^{\pm}_{m+n}&
 &\quad &
 \bigl[K^{3}_{m},K^{3}_{n}\bigr] 
={}&
 \frac{k}{2}m\delta_{m,-n}
 \nonumber\\
 \bigl[K^{\pm}_{m},\chi^{3}_{r}\bigr]
 ={}&
 \mp\chi^{\pm}_{m+r}&
 &\quad &
 \bigl[K^{3}_{m},\chi^{\pm}_{r}\bigr]
 ={}&
 \pm\chi^{\pm}_{m+r}&
 &\quad &
 \bigl[K^{\pm}_{m},\chi^{\mp}_{r}\bigr]
 ={}&
 \pm 2\chi^{3}_{m+r}
 \\
 \bigl\{\chi^{+}_{r},\chi^{-}_{s}\bigr\}
 ={}&
 k\delta_{r,-s}&
 &\quad &
 \bigl\{\chi^{3}_{r},\chi^{3}_{s}\bigr\}
 ={}&
 \frac{k}{2}\delta_{r,-s}\,.&
 \nonumber
\end{alignat}

\noindent The dual Coxeter number is $h^\vee_{\sutwo} = +2\,$. As for the case of $\sltwof$, we can decouple the bosons from the femions
by defining 
\begin{equation}\label{decoupled K currents 1}
\mathcal{K}^{3} =  K^{3} -\frac{1}{k}\left(\chi^{+}\chi^{-}\right) \ , \qquad \mathcal{K}^{\pm} =  K^{\pm} \mp \frac{2}{k}\left(\chi^{3}\chi^{\pm}\right)\ ,
% \label{decoupled K currents 2}
\end{equation}
so that $\bigl[\cK^{a}_{m},\chi^{b}_{n}\bigr] =0$. The decoupled currents satisfy again the same algebra as the $K^{a}$, but 
with level $(k-2)$ instead. We will therefore mostly restrict ourselves to $k \geq 2$ in this paper, 
see however the discussion in Section~\ref{sec:5.1.1}.\footnote{It is potentially interesting to study 
the model for certain smaller values of $k$ such as $k=0\,$, see e.g. \cite{Lindstrom:2003mg,Bakas:2004jq} and 
references therein for attempts in this direction (in bosonic setups). While the $k=0$ worldsheet theory is not a 
standard CFT, it may be related to an integrable theory such as the principal chiral model \cite{Bershadsky:1999hk,Gotz:2006qp}.}

The ground states of the corresponding WZW models will transform in the same representation for left- and right-movers 
with respect to the decoupled $\mathfrak{su}(2)$ algebras (i.e., with respect to the zero modes of 
(\ref{decoupled K currents 1})). These representations are labeled by a spin $j'$ with $j' = 0,\tfrac{1}{2},1,\tfrac{3}{2},\ldots$,
and their states are described by $m'=-j,-j'+1\,,\ldots ,j'-1,j'$, as is well-known for $\mathfrak{su}(2)$ representations. 
We choose the convention that the Casimir of the global decoupled algebra (i.e., of the zero modes of \eqref{decoupled K currents 1}) 
on the representation $j'$ equals 
\begin{equation}
\mathcal{C}^{\sutwo}_{2}\vert j',m'\rangle_{{\rm S}^{3}} = j'(j'+1)\vert j' ,m'\rangle_{{\rm S}^{3}}  \ . 
\end{equation}

The decoupled and coupled bosonic zero modes agree in the 
NS-sector, while in the R-sector they differ by a fermionic contribution, and as a consequence, the $K^3_0$ eigenvalues
in the R-sector are shifted by $\pm \frac{1}{2}$ relative to those in the NS-sector, cf., the discussion around eq.~(\ref{J0shift}) above. 

Finally, the $\mathds{T}^{4}$ theory corresponds to four free bosons $Y^{i}$ and four free fermions $\lambda^{i}$ ($i=1,2,3,4$). The ground
states in this sector are characterised by a momentum vector $\vert \vec{p}\,\rangle$ with
\begin{equation}
\left(\partial Y^{i}\right)_{0}\vert \vec{p}\,\rangle =  p^{i}\vert \vec{p}\,\rangle \qquad \text{and}\qquad 
 L_{0}^{\mathds{T}^{4}} \vert \vec{p}\,\rangle = \frac{1}{2}\sum_{i=1}^{4}\left(p^{i}\right)^{2}\vert \vec{p}\,\rangle\ .
\end{equation}
For a compact torus the left- and right-moving momenta need not agree --- they can differ by winding numbers. 
However, for our purposes, i.e., for identifying the leading Regge trajectory, we will always work in the 
zero momentum sector $\vec{p}=\vec{0}$, both for left- and right-movers. The multiplicity of the Ramond sector 
ground states is accounted for as usual by introducing two labels ($s_{2}, s_{3}$), with $s_{2,3}=\pm$.  

%%%%%%%%%%%%%%%%%%%%%%%%%%%%%%%%%%%
\subsection{GSO projection}
%%%%%%%%%%%%%%%%%%%%%%%%%%%%%%%%%%%
As usual in a NS-R worldsheet string theory, one must impose an appropriate GSO projection in order to remove tachyonic modes and guarantee supersymmetry of the spacetime theory. In the NS sector the worldsheet parity operator is defined to be odd on the ground states
\begin{equation}
(-1)^F\vert 0\rangle_{\text{NS}} =  -\vert 0\rangle_{\text{NS}} \ . 
\end{equation}
Let us denote by $N$ the (integer or half-integer) excitation number in the $\mathfrak{sl}(2,\mathds{R})$ sector, while $N'$ is the corresponding
number for $\mathfrak{su}(2)$, and $N''$ for the  $\mathds{T}^4$ excitations. On a state with excitation numbers $(N,N',N'')$
the total worldsheet parity is then 
\begin{equation}
(-1)^F  = -(-1)^{2N+2N'+2N''} \ .
\end{equation}
The GSO projection $(-1)^F=(-1)^{\bar F}=+1$ in the NS-sector thus requires that either one or all three excitation
numbers are half-integer, and this has to be imposed both for left- and right-movers. In order to describe this compactly we introduce the 
number 
\begin{equation}\label{definition n}
n \equiv N +N' +N'' -\nu\ , \qquad \text{where}\qquad \nu  =
 \left\{
\begin{array}{rl}
\displaystyle{\tfrac{1}{2}} & \text{ NS sector} \\ 
 0 & \text{ R sector}
\end{array} 
\right.
\end{equation}
The above considerations imply that $n$ has to be an integer in the NS sector, both for left- and right-movers. Obviously,
the same is true in the R sector since there all excitation numbers are integers anyway.

In the R sector, the GSO projection involves also a contribution from the fermionic zero modes
corresponding to $s_0  s_1  s_2  s_3$. Thus we can, for any descendant, satisfy the GSO 
projection by changing $s_3$, if necessary. Thus the GSO projection is correctly accounted for
by reducing the multiplicity of the $4$-fold ground state in the R-sector of $\mathds{T}^4$ --- corresponding to the 
four choices for $(s_2,s_3)$ with $s_{2,3}=\pm$ ---  to $2$.

%%%%%%%%%%%%%%%%%%%%%%%%%%%%%%%%%%%
\subsection{Physical state conditions}\label{sec:physstate}
%%%%%%%%%%%%%%%%%%%%%%%%%%%%%%%%%%%

The $\mathfrak{sl}(2,\mathds{R})$ WZW model contains a time-like direction, and as a consequence the theory is non-unitary. 
As usual in worldsheet string theory, the corresponding negative-norm states are removed upon imposing the Virasoro constraints. 
In our context, the physical state conditions are
\begin{align}\label{Virasoro constraints}
L^{\text{tot}}_0-\nu =\bar{L}^{\text{tot}}_0-\bar{\nu} = 0 \  ,
\end{align}
where $\nu,\bar{\nu} =0,\frac{1}{2}$ in the R and NS sectors, respectively, and 
$L^{\text{tot}}_{0} =  L_{0}^{\sltwo}+ L_{0}^{\sutwo} +  L_{0}^{\mathds{T}^4}$. 
We parameterise the contributions from each component as
\begin{alignat}{4}
 L_{0}^{\sltwo}
  ={}&
   -\frac{j(j-1)}{k} + N &
   &\qquad &
 \bar{L}_{0}^{\sltwo} 
 ={}&
  -\frac{j(j-1)}{k} + \bar{N}
  \\
  L_{0}^{\sutwo} 
  ={}& \frac{j'(j'+1)}{k} + N'&
   &\qquad  &
 \bar{L}_{0}^{\sutwo} 
 ={}&
  \frac{j'(j'+1)}{k} + \bar{N}' 
 \\
  L_{0}^{\mathds{T}^4}
   ={}& h^{\mathds{T}}+ N'' &
  &  \qquad&
 \bar{L}_{0}^{\mathds{T}^{4}}
   ={}&
    h^{\mathds{T}}+ \bar{N}'' \ .
\end{alignat}
Here, $j$, $j'$ and $h^{\mathds{T}}$ label the spins (resp.\ the conformal dimension) of the corresponding ground states;
for the case of $\mathfrak{sl}(2,\mathds{R})$ and $\mathfrak{su}(2)$ the relevant spins are defined with respect to the
decoupled currents. 
Furthermore, physical states satisfy the super-Virasoro constraints
\begin{alignat}{3}
 L^{\text{tot}}_{m}\vert \text{phys} \rangle & = 0 & \qquad & m > 0\\
 G^{\text{tot}}_r\vert \text{phys} \rangle & = 0 & \qquad & r>0 \, ,
\end{alignat}
where again $L^{\text{tot}}$ and $G^{\text{tot}}$ denote the total worldsheet currents, receiving contributions from all three sectors of the theory. 

The no-ghost theorem \cite{Balog:1988jb,Petropoulos:1989fc,Dixon:1989cg,Hwang:1990aq,Evans:1998qu} (adapted here to the 
supersymmetric setup, see  also \cite{Pakman:2003cu}) shows that the Virasoro constraints \eqref{Virasoro constraints} remove 
negative-norm states from the spectrum 
provided the unitarity bound 
\begin{equation}\label{unitarity bound}
0 \leq j \leq \frac{k+2}{2}
\end{equation}
is satisfied. This condition is the $k$-dependent bound on the spin $j$ that we mentioned before, see the discussion at the end of 
Section~\ref{sec:AdS3WZW}. It was argued in \cite{Maldacena:2000hw}, based on the structure of the spectrally flowed 
representations, that in fact the bound on $j$ should be slightly stronger and take the form 
\begin{equation}\label{MO bound}
\frac{1}{2} < j < \frac{k+1}{2} \ . 
\end{equation}
For most of the following the (weaker) unitarity bound will suffice, but for some arguments, in particular
the analysis of the spectrally flowed representations, the stronger  Maldacena-Ooguri (MO) bound (\ref{MO bound})
will be required. 

Next, we write the first equation in the on-shell condition \eqref{Virasoro constraints} as 
\begin{equation}\label{on-shell condition v2}
-\frac{j(j-1)}{k} + \frac{j'(j'+1)}{k} + h^{\mathds{T}} + n = 0\ , 
\end{equation}
where $n$ was defined above in eq.~(\ref{definition n}). In addition, we get the same equation with $\bar{n}$ in place of $n$ from
the second condition of  \eqref{Virasoro constraints}, where $\bar{n}$ is defined analogously for the right-movers. We therefore conclude
that $n=\bar{n}$. Furthermore, as was noted above, $n$ is always a non-negative integer after GSO-projection. We can use eq.~(\ref{on-shell condition v2}) to solve for $j$ as\footnote{We have taken here the positive square root since $j>0$ for unitarity.}
\begin{equation}\label{jprime}
j =    \frac{1}{2} \left( 1 + \sqrt{ \left(2j'+1\right)^2+4k \left(n + h^{\mathds{T}} \right)}\, \right) \ . 
%\frac{1}{2} + \sqrt{ \left(j'+\frac{1}{2}\right)^2+k n}\, .
\end{equation}
Note that for fixed $n$,  the Virasoro level of the physical states satisfy $N\,,\, N' \,,\, N'' \leq n + \nu$, 
as follows from \eqref{definition n}, and similarly in the barred sector. Since each excitation mode can raise the $J_0^3$ eigenvalue
at most by one (and since each fermionic $\psi^{\pm}_{-1/2}$ mode can only be applied at most once), we conclude that in the NS sector 
the $J_{0}^{3}$ eigenvalue $m$ of the physical states will lie between $j-n-1 \leq m \leq j+n+1$, while in the R sector it will lie 
between $j-n-1/2 \leq m \leq j+n +1/2$. This implies 
that the spacetime states labeled by $n$ have spin $s$ bounded  as $|s| \leq 2n+2$. More explicitly, 
the relevant states are of the form\footnote{From what we have said so far, it is not yet clear that all these states will
indeed be physical, but this will turn out to be the case, see the discussion below in Section~\ref{sec: Regge}. Furthermore, 
some of these states will appear with higher multiplicity. For the arguments of the next section it is however enough to 
know that only these charges can appear among the physical states.}
\begin{gather}\label{general discrete state}
\vert j +r-n-1 \rangle \otimes  \overline{\vert j +\bar{r} -n-1 \rangle}\,,\qquad \text{with}\quad 0 \leq r, \bar{r} \leq 2n+2 \ , 
\end{gather}
where $r$ and $\bar{r}$ are positive integers or zero in the NS sector, and positive half-integers in the R sector --- these
parameters are simply related to $(\ell,\bar{\ell})$, see the paragraph above (\ref{ellell}), 
by a shift in order to make them non-negative. 
The spacetime energy and spin of these states is then given by 
\begin{equation}\label{general energy and spin}
E = 2j + r  +\bar{r} - 2n-2\,,\qquad s = r - \bar{r}\ ,
\end{equation}
which in particular implies
\begin{equation}\label{dispersion relation in terms of spin}
E = s+2\left(j  +\bar{r} - n-1\right)\ .
\end{equation}

Finally, it is worth pointing out how the AdS$_{3}\times {\rm S}^{3} \times \mathds{T}^{4}$ supergravity spectrum is obtained from the worldsheet 
description. The supergravity states all arise for $n=0$, which leads then to fields of 
spin $\vert s \vert \leq 2$.
Furthermore, this condition restricts the excitation levels as $N,N',N'' \leq \nu$, and it follows that the supergravity spectrum is 
obtained from the level $1/2$ descendants in the NS sector, as well as the R ground states. 
Crucially, from \eqref{jprime} (with no momentum in the $\mathds{T}^{4}$ directions) we deduce that the $\sltwo$ and $\sutwo$ spins are related by
\begin{equation}
\text{SUGRA:} \qquad 
j = j'+1\ ,
\end{equation}
so that $j$ is now an integer or half integer (with $j \geq 1$). We have explicitly checked that the corresponding physical states precisely 
reproduce the supergravity spectrum, as derived in e.g.\ \cite{deBoer:1998ip}. In particular, one finds that the $j'=0$ ($j=1$) sector 
contains the (massless) graviton supermultiplet, while the representations with $j' > 0$ give rise to a tower of massive BPS 
multiplets.\footnote{Indeed, it is easy to see from \eqref{dispersion relation in terms of spin} that chiral ($E=s$) states in 
the $n=0$ sector can only exist for $j=1$ (and $\bar{r}=0$).}

%%%%%%%%%%%%%%%%%%%%%%%%%%%%%%%%%%%%%%%%%%%%%%%%%%%%%%%%%%%%%%%%%%%%
\section{No massless higher spin states from short strings}\label{sec:no hs}
%%%%%%%%%%%%%%%%%%%%%%%%%%%%%%%%%%%%%%%%%%%%%%%%%%%%%%%%%%%%%%%%%%%%

With these preparations at hand we now want to analyse whether the string spectrum possesses massless higher spin states
at least for some value of the level. As we shall show in this section, this will not be the case for the short strings
coming from the unflowed (discrete) representations. 

Recall first the standard holographic relation between the mass of an AdS$_{3}$ (bulk) excitation and the conformal dimension $E$ and 
spin $s$ of the dual operator in the boundary 2$d$ CFT \cite{Aharony:1999ti}, 
\begin{equation}
m_{\text{bulk}}^{2} = \bigl(E-|s|\bigr)\bigl(E +|s|-2\bigr)\ , 
\end{equation}
where $E=h+\bar{h}$ and $s = h-\bar{h}$ in the usual CFT notation. As expected, massless higher spin fields are dual to conserved 
currents of dimension greater than two, which in the present context satisfy $E=|s|$ (with $|s| > 2$). Hence, massless higher spin states
are characterised by the property that either the $J_{0}^{3}$ eigenvalue or the $\bar{J}_{0}^{3}$ eigenvalue vanishes.

Let us concentrate, for concreteness, on the case $\bar{J}_{0}^{3}=0$. Then it follows from (\ref{general discrete state}) that we need to have 
\begin{equation}\label{chiral j}
\text{conserved current}\qquad \Rightarrow \qquad j = n + 1 -\bar{r}\ . 
\end{equation}
Then the on-shell condition \eqref{on-shell condition v2} implies that 
\begin{equation}\label{chiral k}
k = \frac{\left(n-\bar{r}-j'\right)\bigl(n+1-\bar{r}+j'\bigr)}{n+h^{\mathds{T}}}\ .
\end{equation}
As discussed above, the case $n=0$ corresponds to (supergravity) states that have $|s|\leq 2$ and are therefore not of higher spin. 
We may therefore assume that $n \geq 1$. Our strategy will be to show that unitarity implies  that 
$n + \bar{r} \leq 1$, contradicting the $n \geq 1$ assumption, except for $n=1$ and $\bar{r}=0$. The latter case is then excluded
by the stronger MO-bound (or by noticing that the relevant state is null). 

First, from \eqref{chiral j} we note that the unitarity bound $j \geq 0$ implies $n + 1 -\bar{r}\geq0$. Since $j' \geq 0$ by definition, 
from \eqref{chiral k} we find that positivity of $k$ requires
\begin{equation}\label{n-barq}
\text{conserved current } +\, (k >0) \qquad \Longrightarrow \qquad n-\bar{r} > j' \geq 0 \ . 
\end{equation}
Next, we use the unitarity bound (\ref{unitarity bound})  which translates for $j = n + 1 -\bar{r}$ into
\begin{equation}
  n- \bar{r} \leq \frac{\bigl(n-\bar{r}-j'\bigr) \bigl(n+1-\bar{r}+j'\bigr)}{2 (n + h^{\mathds{T}})}\ .
\end{equation}
Together with \eqref{n-barq}, this requirement is equivalent to
\begin{equation}\label{3.6}
 h^{\mathds{T}}+\frac{j'(j'+1)}{2(n-\bar{r})}\leq \frac{1-n-\bar{r}}{2}\ .
\end{equation} 
Since the quantity on the left hand side is greater or equal to zero (recall that $n -\bar{r} >0$ and $h^{\mathds{T}}\geq 0$, 
$j' \geq 0$ by unitarity), we conclude $n+\bar{r} \leq 1$. Finally, for $n=1$ and $\bar{r}=0$, we have $j=2$, and hence from 
(\ref{jprime}), $k\leq 2$, which is only compatible with unitarity for $k=2$ (and incompatible with the stronger MO-bound 
(\ref{MO bound}) even in that case).  Actually, the corresponding state 
\begin{equation}
\bar{{\cal J}}^{-}_{-1} \bar{\psi}^{-}_{-1/2} \,\overline{ \left|j=2\right\rangle }
\end{equation}
is null at $k=2$, as has to be the case since it saturates the unitarity bound. 
 
Summarizing, we have shown that the only conserved currents that exist in the unflowed discrete representations appear in the supergravity
spectrum $(n=0)$, and thus have spin $s\leq 2$. 
Our analysis holds for all values of the level $k>0$; thus, among the WZW backgrounds
there is no radius at which the theory develops a higher spin symmetry from the short string spectrum. 
This is in line with the arguments of the Introduction, see eq.~(\ref{Tensioneq}). It is also in accord with the results of 
\cite{Gaberdiel:2015uca} where evidence was found  that the symmetric
orbifold point (that exhibits a large higher spin symmetry) is dual to a background with R-R flux. 

The long string sector (that is described by spectrally flowed representations) will be discussed in Section~\ref{sec: spectral}. As 
we shall explain there, for $k=1$ a stringy tower of higher spin fields appears from the spectrally flowed continuous representation, 
mirroring the bosonic analysis of \cite{GGH}. Since these massless higher spin fields arise from long strings, they describe
a qualitatively different higher spin symmetry from the usual tensionless limit \cite{GGH}.

%%%%%%%%%%%%%%%%%%%%%%%%%%%%%%%%%%%%%%%%%%%%%%%%%%%%%%%%%%%%%%%%%%%%
\section{Regge trajectories and their $\cN=4$ structure}\label{sec: Regge}
%%%%%%%%%%%%%%%%%%%%%%%%%%%%%%%%%%%%%%%%%%%%%%%%%%%%%%%%%%%%%%%%%%%%

Next we want to identify the leading Regge trajectory states in the string spectrum and compare this to the 
${\cal W}_\infty$ symmetry that was found in \cite{Gaberdiel:2014cha}. 
In order to identify the leading (and sub-leading) Regge trajectory states in the string spectrum, we first need to study in more detail 
the actual physical states. In this section we concentrate again on the states from the unflowed discrete representations;
the spectrally flowed representations will be discussed in Section~\ref{sec: spectral}.

%%%%%%%%%%%%%%%%%%%%%%%%%%%%%%%%%%%
\subsection{General discrete spectrum}
%%%%%%%%%%%%%%%%%%%%%%%%%%%%%%%%%%%

Recall from our discussion in Section~\ref{sec:physstate} 
that physical states in a representation built from an AdS$_{3}$ groundstate labeled by $j$ take the 
form \eqref{general discrete state}, with the corresponding spacetime energy and spin being given by \eqref{general energy and spin}. 
We now want to show that for all choices of $r,\bar{r}$ in $0 \leq r,\bar{r} \leq 2n+2$, physical states with these 
quantum numbers exist. In addition, we want to determine their multiplicities.

Let us start with some general comments about the string spectrum. 
One should expect that the physical states are obtained by applying eight transverse oscillators to the ground states --- of
the ten oscillators, one linear combination is eliminated by the Virasoro condition, and a second one leads to spurious states, i.e.,
gauge degrees of freedom. In the current context, it is natural to take the light-cone directions to be a linear combination of 
the time-like AdS$_3$ direction, as well as one direction on the $\mathds{T}^4$. Then the transverse (physical) excitations 
correspond to the $\pm$ modes from AdS$_3$, all three oscillators from the ${\rm S}^3$ factor, and three of the four oscillators 
from the $\mathds{T}^4$. Thus the physical descendants of the ground states of the chiral NS and R sector are expected 
to be counted by  --- here $j$ and $j'$ label the spins of the $\mathfrak{sl}(2,\mathds{R})$ and $\mathfrak{su}(2)$ ground
state representation (taken with respect to the decoupled currents), respectively,\footnote{As far as we are aware, this 
formula was first written down in  \cite{Raju:2007uj} generalizing the corresponding bosonic formula from \cite{Maldacena:2000kv}
and building on \cite{Israel:2003ry}. These formulae are correct for sufficiently large values of $k$ for which there are no 
non-trivial null-vectors.}
\begin{align}\label{characters}
\chi^{\text{NS}}(q,z,y) 
={}&
 q^{h(j)+h'(j')+h^{\mathds{T}}} \frac{y^{j}}{1-y}\,  \frac{(z^{j'+1} - z^{-j'})}{(z-1)} \\
& \hspace*{-1.5cm}
 \times \prod_{n=1}^{\infty} \frac{(1+yq^{n-1/2})(1+y^{-1}q^{n-1/2})(1+zq^{n-1/2})(1+z^{-1}q^{n-1/2})(1+q^{n-1/2})^4}
 {(1-yq^n)(1-y^{-1}q^n)(1-zq^n)(1-z^{-1}q^n)(1-q^n)^4}
    \nonumber \\
&    \hphantom{a}
    \nonumber\\
%\end{align}
%\begin{align}
\chi^{\text{R}}(q,z,y) 
={}&
 2 q^{h(j)+h'(j')+h^{\mathds{T}}}\, \frac{y^j\, \bigl(y^{1/2}+y^{-1/2}\bigr)}{(1-y)}  \frac{(z^{j'+1} - z^{-j'})(z^{1/2}+z^{-1/2})}{(z-1)}
\label{characters 2}
\\
&
\times \prod_{n=1}^{\infty}\frac{(1+yq^n)(1+y^{-1}q^n)(1+zq^n)(1+z^{-1}q^n)(1+q^n)^4}
{(1-yq^n)(1-y^{-1}q^n)(1-zq^n)(1-z^{-1}q^n)(1-q^n)^4} 
\nonumber
\end{align}
where $h^{\mathds{T}}$ is the ground state conformal dimension of the $\mathds{T}^4$ theory, while for the 
$\mathfrak{sl}(2,\mathds{R})$ and $\mathfrak{su}(2)$ factors we have
\begin{align}
h(j)  = -\frac{j(j-1)}{k}  \ , \qquad h'(j')  =\frac{j'(j'+1)}{k} \ .
\end{align}
Here $y$ and $z$ are the chemical potentials with respect to $\mathfrak{sl}(2,\mathds{R})$ and $\mathfrak{su}(2)$,
respectively, and we have used that the corresponding characters are of the form
\begin{equation}\label{sl2char}
\chi_j(y) = \frac{y^j}{1-y} \ , \qquad \chi_{j'}(z) = \sum_{k=-j'}^{j'}z^{k} = \frac{(z^{j'+1} - z^{-j'})}{(z-1)}  \ .
\end{equation}
Furthermore, $q$ keeps track of the total Virasoro eigenvalue which has to equal
$q^{\nu}$ for the actual physical states, see eq.~(\ref{Virasoro constraints}). (We are here describing one chiral sector; 
the results for left- and right-movers then has to be combined.) 
The first line in each of \eqref{characters}-\eqref{characters 2} accounts for the contribution of the ground state representations, 
while the second line describes 
the contributions of the non-zero oscillators. The overall multiplicity of $2$ in the R-sector reflects the overall multiplicity after
GSO projection, see the discussion after eq.~(\ref{definition n}). 

We have checked this prediction in some detail (by solving the physical state conditions explicitly, at least for some
low-lying states), and we have found complete agreement. We should mention, though, that there
are some subtleties with the counting for $j=1$; this is discussed in more detail in 
Appendix~\ref{app:physstatessubtleties}. 

We note that this formula in particular implies that, for all $0 \leq r \leq 2n+2$, physical states with 
these quantum numbers exist. In order to see this, we solve for $j$ (in terms of $n$, $j'$ and $h^{\mathds{T}}$) using
eq.~(\ref{jprime}); then the overall power of $q^{\nu}$ comes from taking the term with $q^{n+\nu}$ from
the oscillator product in the second line. In the NS sector $r=0$ then corresponds to the situation where 
the $J^3_0$ eigenvalue is $j-n-1$.  This can be achieved by taking from the numerator the term $y^{-1}q^{1/2}$,
as well as $n$ powers of $y^{-1}q^1$ from the geometric series expansion of the denominator term $(1-y^{-1} q)$. The 
corresponding state is thus of the form 
\begin{equation}\label{right moving extremal}
\vert j-n-1\rangle=
 \bigl(\mathcal{J}_{-1}^{-}\bigr)^{n}\psi_{-1/2}^{-}\vert j\rangle\ .
\end{equation}
Similarly, the case $r=2n+2$ corresponds to having $J^3_0$ eigenvalue $j+n+1$, in which case the relevant
powers are $yq^{1/2}$ from the numerator, and $n$ powers of $yq^1$ from the geometric series expansion of the 
denominator term $(1-y q)$. Schematically, the corresponding state is thus of the form
\begin{equation}\label{left moving extremal}
\vert j+n+1\rangle =\left[\bigl(\mathcal{J}_{-1}^{+}\bigr)^{n}\psi_{-1/2}^+\vert j\rangle  +\cdots\right]\ ,
\end{equation}
where the dots stand for additional terms that make it a lowest weight state with respect to the 
$\mathfrak{sl}(2,\mathds{R})$ algebra. In either case it is easy to see that these 
representations appear with multiplicity one --- these are the `extremal' cases that can only be obtained in one way. 

On the other hand, the intermediate cases $0 < r < 2n+2$ can be obtained in more than one way, but
from the above analysis it is clear that all of these terms will indeed arise. Incidentally, we should note that it
follows from the explicit formula that (apart from the overall $y^j/(1-y)$ term) the partition function is
symmetric under the symmetry $y\leftrightarrow y^{-1}$. As a consequence, the multiplicities of the
representations corresponding to $r$ and $2n+2-r$ will be the same.

\begin{figure}[h!]
\centering
\includegraphics[width=15cm]{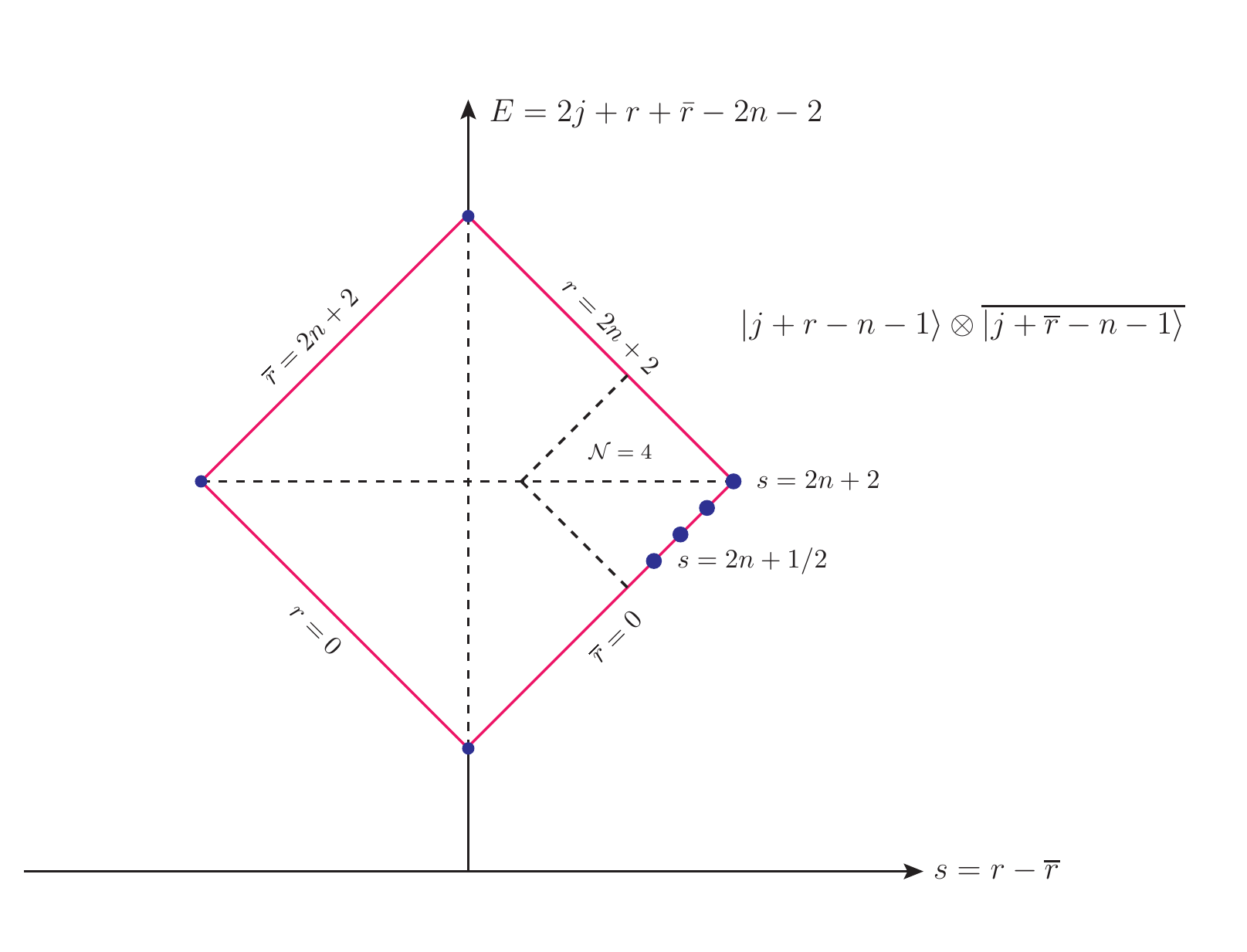}
\caption{Schematic description of the discrete physical spectrum for fixed $n$ (and fixed $j$). The four blue dots on the $\bar{r}=0$ edge denote the top component of supermultiplets with the lowest energy for the given spin. See section \ref{sec:LR} below for a detailed discussion of these states.}
\label{fig:diamond1}
\end{figure}

Combining left- and right-movers, the full spacetime spectrum (in terms of energy and spin) 
forms a diamond in the $(E,s)$ plane for fixed $n$, depicted in Figure~\ref{fig:diamond1}. Here the corner
points have multiplicity one, but the other points have higher multiplicity. 

On general grounds it is clear that we must be able to organise the spectrum in terms of 
(small) $\mathcal{N}=(4,4)$ representations, see Appendix~\ref{app: multiplets} for a brief
review of their structure. In the $(E,s)$ plane, $\mathcal{N}=(4,4)$ multiplets form small diamonds with 
edges spanning two units of energy and two units of spin.  For example, the right most vertex
of the diamond in Figure~\ref{fig:diamond1}, which is characterised by $(r,\bar{r})=(2n+2,0)$,
has multiplicity one (since both $r$ and $\bar{r}$ take their extremal values), and corresponds
to the chiral states with $h=j+n+1$ and $\bar{h}=j-n-1$. This state is then the top ($h=h_0+2$) 
component of the left-moving  long ${\cal N}=4$ multiplet whose bottom component has $h_0=j+n-1$ and 
transforms in the representation ${\bf m}$ of $\mathfrak{su}(2)$, where $m=2j'+1$, see 
Table~\ref{table: chiral long multiplet} of Appendix~\ref{app: multiplets}. 
Similarly, with respect to the right-movers, the state is the bottom component of 
a similar ${\cal N}=4$ multiplet with $\bar{h}_0 = j-n-1$. The relevant states in the full
multiplet then give rise to states in the dashed diamond in Figure~\ref{fig:diamond1}. (Here we have also
included the R sector states that are needed to complete the multiplets.)

Once the states that sit in this multiplet have been accounted for, we look at the remaining states
and proceed iteratively. For example, the `extremal' R sector states that contribute to this multiplet
have  $h=j+n+\frac{1}{2}$ and/or $\bar{h} = j - n - \frac{1}{2}$. Concentrating on the first case,
it follows from (\ref{characters 2}) that there will be $8 m$ states of this form transforming  as
$4\cdot ({\bf m}+1)$  and $4\cdot ({\bf m}-1)$ --- one factor of $2$ is the overall factor in eq.~(\ref{characters 2}), 
while the other factor of $2$ comes from the fact that we can either use one fermionic $(-1)$ mode in the R-sector or none.
Furthermore,  the two different representations come from tensoring with the spin $\frac{1}{2}$ representation
described by the factor $(z^{1/2} + z^{-1/2})$ in the first line. 
Two copies of each of these two representations are part of the long ${\cal N}=4$ multiplet, see Table~\ref{table: chiral long multiplet},
while the other two will generate two pairs of new ${\cal N}=(4,4)$ multiplets, whose
bottom components will transform as $({\bf m}+1)$ and $({\bf m}-1)$, respectively. 
(The second dot along the $\bar{r}=0$ edge in Figure~\ref{fig:diamond1} represents states in these multiplets.)
Proceeding in this manner, we find that the multiplicity of the ${\cal N}=4$ multiplets along the $\bar{r}=0$ edge 
(i.e., only considering states whose bottom component is $\bar{h}_0 = j-n-1$) is described in 
Table~\ref{table: N = 4 multiplets}.

\begin{table}[h!]
\centering
\begin{tabular}{c|ccccccccccc}
$s$ & $\delta m=-5$&${-4}$&$-3$&$-2$&$-1$& $0$ & $1$&$2$&$3$&$4$&$5$ \\
\hline
$2n+2$   &  &   &    &    &     & 1  &     &    &    &   & \\
$2n+3/2$ &  &   &    &    & 2   &    & 2   &    &    &   & \\
$2n+1$   &  &   &    & 2  &     & 9  &     & 2  &    &   & \\
$2n+1/2$ &  &   & 2  &    & 18  &    & 18  &    & 2  &   & \\
$2n$     &  & 2 &    & 23 &     & 61 &     & 23 &    & 2 & \\
$2n-1/2$ &2 &   & 24 &    & 116 &    & 116 &    & 24 &   & 2 \\
$\ldots$ &$\ldots$ & $\ldots$  & $\ldots$ & $\ldots$   & $\ldots$ & $\ldots$   & $\ldots$ & $\ldots$ & $\ldots$ & $\ldots$  & $\ldots$ \\
\end{tabular}
\caption{Multiplicity of the $\mathcal{N}=4$ multiplets that arise along the $\bar{r}=0$ edge for the case
where $j'$ and hence ${\bf m}$ is sufficiently big. Here and below the spin always refers to the spin $s$ of the `top'
component of the ${\cal N}=4$ multiplet --- the bottom component then has spin $s-2$.
}
\label{table: N = 4 multiplets}
\end{table}

\noindent For future reference,  in Table~\ref{table: jp = 0 multiplets} we also give the multiplicity of the ${\cal N}=4$ 
multiplets along the $\bar{r}=0$ edge for $j'=0$, i.e., ${\bf m}=1$ --- in this case, only $\delta m\geq 0$ 
is possible and some of the multiplicities are reduced. 
\begin{table}[h!]
\centering
\begin{tabular}{c|ccccccc}
         &${\bf 1}$ & ${\bf 2}$ & ${\bf 3}$ & ${\bf 4}$ & ${\bf 5}$ & ${\bf 6}$ &\ldots \\ 
\hline
$2n+2$   & 1   &     &    &    &    & &\\
$2n+3/2$ &     &  2  &    &    &    & &\\
$2n+1$   & 7   &     & 2  &    &    & &\\
$2n+1/2$ &     & 16  &    & 2  &    & & \\
$2n$     & 38  & 	  & 21 &    &  2 &  & \\
$2n-1/2$ &     &  92 &    & 22 &    & 2 & \\
\ldots & \ldots & \ldots & \ldots & \ldots & \ldots & \ldots    
\end{tabular}
\caption{Multiplicities of $\mathcal{N}=4$ multiplets along the $\bar{r}=0$ edge for $j'=0$, i.e., ${\bf m}=1$.}
\label{table: jp = 0 multiplets}
\end{table}

%%%%%%%%%%%%%%%%%%%%%%%%%%%%%%%%%%%
\subsection{Leading Regge trajectory}\label{sec:LR}
%%%%%%%%%%%%%%%%%%%%%%%%%%%%%%%%%%%

Having discussed the general structure of the discrete string spectrum, we can now identify
the states on the leading Regge trajectory. These are the states that should have the lowest
energy for a given spin, together with their ${\cal N}=(4,4)$ descendants. We want to argue
that they are precisely described by the dashed diamond in Figure~\ref{fig:diamond1}, where
$n$ takes the values $n=0$, $n=1$, $n=2$, etc. 

First we note that the leading Regge trajectory states will be associated to states with $j'=0$
(and $h^{\mathds{T}}=0$) --- for fixed $n$, as well as $(r,\bar{r})$, the choice of $j'$ and $h^{\mathds{T}}$ only enters via
$j$ as defined in (\ref{jprime}), and $j$ in turn only contributes to $E$, but not to $s$, see eq.~(\ref{general energy and spin}). 
Choosing $j'$ and or $h^{\mathds{T}}$ to be non-trivial, increases $j$ and hence $E$, but does not modify the spin $s$. 
The states of lowest energy (for fixed spin) therefore arise for $j'=h^{\mathds{T}}=0$. 

\begin{figure}[h]
\centering
\includegraphics[width=15cm]{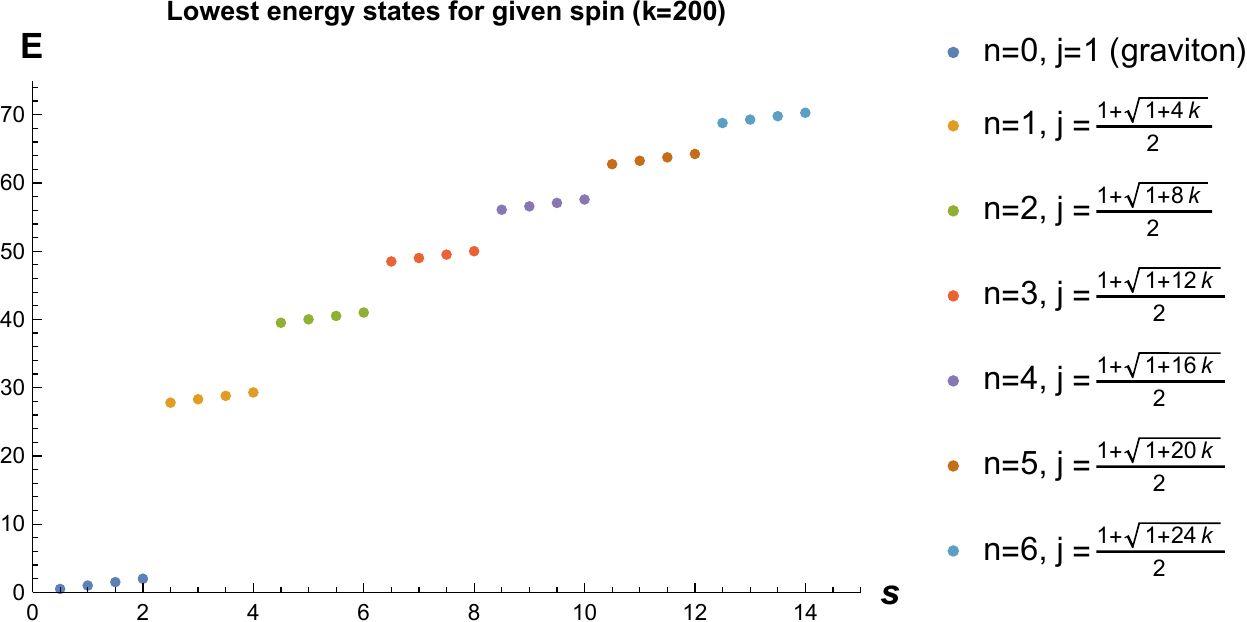}
\caption{Dispersion relation of the lowest energy states for $k=200\,$. The dots correspond to the set of four states singled out in 
Figure \ref{fig:diamond1}, depicted here for different values of $n$ (i.e. different $j$), giving rise to different ranges $2n +1/2 \leq s \leq 2n+2$ 
of the spin. In each family of four dots, the last one describes a unique state, the corner state of Figure \ref{fig:diamond1}, while the 
first three correspond to states with higher multiplicity.}
\label{fig:Regge200}
\end{figure}

Similarly, by construction, the states with lowest energy for given spin lie (for fixed $n$ and hence $j$ --- recall
that $j'=h^{\mathds{T}}=0$) 
on the lower edges of the representation diamond. Without loss of generality, focusing on positive helicity 
modes we can then restrict our attention to the $\bar{r}=0$ edge. The energies of these states satisfy 
the linear dispersion relation
\begin{equation}\label{bredge}
E_{\text{Regge}} (s) =s-2n-1+\sqrt{1+4k n}\ , \qquad \qquad  2n <s \leq 2n+2\ ,
\end{equation}
where the inequality $2n < s$ arises because the lowest energy state with spin 
$s=2n$ is obtained from the diamond corresponding to $\tilde{n} = n-1$ --- this is a consequence
of the inequality
\begin{equation}
\sqrt{1+4kn} \geq  2 + \sqrt{1+4k(n-1)} \ , 
\end{equation}
which, after squaring twice, is equivalent to 
\begin{equation}\label{cons1}
(k+2) \geq  4 n \ ; 
\end{equation}
in turn this follows from the unitarity bound, see eq.~(\ref{unitarity bound}), using the expression for $j$ from eq.~(\ref{jprime}) 
with $j'=h^{\mathds{T}}=0$. The conformal dimensions of the leading Regge trajectory states for small  values of the spin are 
plotted (for $k=200$) in Figure~\ref{fig:Regge200}.

As a side remark, we should note that the states with dispersion relation (\ref{bredge}) formally become chiral if $k$
takes the value $k=n+1$. However, this choice is not allowed by the unitarity bound, except for the 
supergravity states with $n=0$ and the special solution $n=1$ that was already discussed after eq.~(\ref{3.6}). [The 
latter case corresponds to $n=1$ and $k=2$ and is incompatible with the MO bound (\ref{MO bound}).] 

Since $\bar{r}=0$ the right-moving states are the `extremal' states with $\bar{N}'=\bar{N}''=0$,
so that the right-moving (barred) $\mathfrak{su}(2)$ representation is always trivial. Furthermore,
the leading term with $r=2n+2$ is also trivial with respect to the left-moving $\mathfrak{su}(2)$
algebra, and it is the top state of an ${\cal N}=(4,4)$ multiplet with $\mathfrak{su}(2) \oplus\mathfrak{su}(2)$
quantum numbers $({\bf 1},{\bf 1})$. 

We now want to argue that the leading Regge trajectory consists just of the first multiplet of Table~\ref{table: jp = 0 multiplets} 
for each $n$. This is natural since there is only a single multiplet with these quantum numbers; its top component is obtained by 
tensoring the $\sltwo$ representations \eqref{right moving extremal} and \eqref{left moving extremal} for the left- and right-moving 
sector, respectively. (The terms with $r<2n+2$, on the other hand, lead in general to ${\cal N}=(4,4)$ multiplets for which the left-moving 
$\mathfrak{su}(2)$ spin is not trivial.) Furthermore, these
states always define the states with smallest energy for the given spin, independent of $k$.\footnote{The situation 
is in general more complicated for the other states, see the discussion of the next subsection.} In order to see this, it is enough to show
that $E(n,s=2n+2) < E(p,s=2n+2)$ for any $p>n$ --- note that a state with this spin can only appear for $p\geq n$. Without loss of 
generality it is enough to concentrate on the case $p=n+1$ since any $p>n$ can be iteratively obtained in this manner. Furthermore,
we may assume that the relevant state in the $p$'th (i.e., $n+1$'th) diamond sits on the lower edge, i.e., has energy described by 
eq.~(\ref{bredge}). Then the inequality we need to prove is simply 
\begin{equation}
\sqrt{1 + 4 k p} \ \geq \ \sqrt{1 + 4 k (p-1)} + 2 \ ,
\end{equation}
which upon squaring both sides (after subtracting $2$) leads to 
\begin{equation}
1 + k \geq  \sqrt{1 + 4 k p}  \ . 
\end{equation}
This identity is now a direct consequence of the unitarity bound, see eq.~(\ref{cons1}) 
with $n=p$.

We note that these states carry exactly the same quantum numbers as the generators of the even spin 
${\cal N}=4$ ${\cal W}_\infty$ algebra that was analysed in \cite{Ferreira:2017zbh}. This is the minimal version of the
${\cal N}=4$ higher spin symmetry, and it has a nice AdS$_3$ dual that is also discussed in some
detail in \cite{Ferreira:2017zbh}. On the other hand, while the string spectrum also contains multiplets with odd
integer spin, there does not seem to be any natural candidate for which of the $7$ singlet
multiplets at spin $2n+1$, see Table~\ref{table: jp = 0 multiplets}, should be added to the even spin
${\cal W}_\infty$ algebra in order to generate the full ${\cal N}=4$ ${\cal W}_\infty$ algebra of 
\cite{Gaberdiel:2014cha} (or the extended algebra of \cite{Gaberdiel:2015uca} where also
the charged bilinears are included in the higher spin algebra). Incidentally, the fact that the leading 
Regge trajectory should only be identified with the fields (or multiplets) of even spin is also expected from bosonic closed
string theory in flat space. There the states of the leading Regge trajectory are associated to the 
worldsheet states of the form 
\begin{equation}\label{flatspace}
\alpha_{-1}^{\mu_1} \cdots \alpha_{-1}^{\mu_n} \, \bar{\alpha}_{-1}^{\nu_1} \cdots \bar{\alpha}_{-1}^{\nu_n}  |p\rangle \ , 
\end{equation}
where the level-matching condition requires that the number of transverse oscillators on the left and right is the same. As a 
consequence, this only leads to fields of even spin $s=2n$.

%%%%%%%%%%%%%%%%%%%%%%%%%%%%%%%%%%%
\subsection{Subleading $\mathcal{N}=4$ trajectory}
%%%%%%%%%%%%%%%%%%%%%%%%%%%%%%%%%%%

Unlike the leading Regge trajectory, the identification of the subleading trajectory turns out to be 
somewhat less clean, and in particular it depends on the value of $k$. For $2n<s\leq 2n+2$ there are a priori three kinds of states competing to be the subleading trajectory. These are 
the states in the interior of the $(n,j'=0)$ diamond; the states on edge of the $(n,j'=1/2)$ diamond; and the 
states on the edge of the $(n+1,j'=0)$ diamond. Denoting the energies of these three sets by $E_{n}^*(j'=0)$, $E_{n}(j'=1/2)$, and $E_{n+1}(j'=0)$, 
respectively, we find that their explicit values for the relevant spins are as given in Table~\ref{Table3}.
\begin{table}[h!]
\centering
\begin{tabular}{|c|c|c|c|}
\hline
s & $E^*_n(j'=0)$ & $E_{n+1}(j'=0)$ & $E_{n}(j'=1/2)$ \\
\hline
$2n+2$   & - & $-1+\sqrt{1+4k(n+1)}$ & $1+2\sqrt{1+kn}$ \\
$2n+3/2$ & $\frac{3}{2}+\sqrt{1+4kn}$ & $-\frac{3}{2}+\sqrt{1+4k(n+1)}$ & $\frac{1}{2}+2\sqrt{1+kn}$ \\
$2n+1$ & $1+\sqrt{1+4kn}$ & $-2+\sqrt{1+4k(n+1)}$ & $2\sqrt{1+kn}$ \\
$2n+1/2$ & $\frac{1}{2}+\sqrt{1+4kn}$ & $-\frac{5}{2}+\sqrt{1+4k(n+1)}$ & $-\frac{1}{2}+2\sqrt{1+kn}$ \\
\hline
\end{tabular}
\caption{Candidates for subleading Regge trajectory with $2n<s\leq 2n+2$.}
\label{Table3}
\end{table}

\noindent It turns out that among these states, the one with the smallest energy is 
\begin{alignat}{4}
E_{n+1}(j'=0) & \qquad & \text{ if } & \qquad & k & \leq k^*_n=\frac{7}{4}+2n+\sqrt{4n^2+7n+4}  \label{case1}\\
E_{n}(j'=1/2) & \qquad & \text{ if } & \qquad & k & \geq k^*_n=\frac{7}{4}+2n+\sqrt{4n^2+7n+4} \; .
\end{alignat}
A few remarks are in order. First, the competing states always lie on the edge of some diamond. Second, 
for fixed $k$, the choice between the two diamonds is $n$- and therefore $s$-dependent. Nevertheless, the existence 
of a minimum value for $n$ (which is $n=0$) implies that we can make the states of eq.~(\ref{case1})  to be the
subleading ones for all possible values of $n$, and thus for all higher spin states, by tuning $k$ to be small enough.
This happens for $k\leq\frac{15}{4}$. Note that since $k$ must be integer, this allows for the two solutions
$k=2$ and $k=3$. 

We should also note that the $\mathfrak{su}(2) \oplus \mathfrak{su}(2)$ quantum numbers are different for these 
two sets of competing representations, as detailed in Table \ref{table: subleading trajectory}. In particular, the states
of the second column are non-trivial with respect to the right-moving $\mathfrak{su}(2)$ algebra. Unfortunately, there 
does not seem to be any particularly simple pattern among these representations, and they do not seem
to be naturally in correspondence with the subleading Regge trajectory of \cite{Gaberdiel:2015uca}.\footnote{We should mention 
that among the above states one should expect that some  do not become chiral at the symmetric orbifold point, i.e., do not belong 
to the stringy higher spin symmetry, but remain massive even at that point in moduli space.}
Obviously, there is no fundamental reason why such a correspondence should exist --- the two descriptions refer to different points in moduli space.  
\begin{table}[h!]
\centering
\begin{tabular}{|c|c|c|}
\hline
s & $k< k_n^*$ & $k>k_n^*$\\
\hline
$2n+2$   & $2\cdot ({\bf 5},{\bf 1})\oplus 21\cdot({\bf 3},{\bf 1}) \oplus 38\cdot({\bf 1},{\bf 1})$ & $ ({\bf 2},{\bf 2})$ \\
$2n+3/2$ & $2\cdot ({\bf 6},{\bf 1})\oplus 26\cdot({\bf 4},{\bf 1}) \oplus 88\cdot({\bf 2},{\bf 1})$ & $2\cdot ({\bf 3},{\bf 2})\oplus 2\cdot ({\bf 1},{\bf 2})$\\
$2n+1$   & $2\cdot ({\bf 7},{\bf 1})\oplus \ldots $ & $2\cdot ({\bf 4},{\bf 2})\oplus 9\cdot ({\bf 2},{\bf 2})$ \\
$2n+1/2$ & $2\cdot ({\bf 8},{\bf 1})\oplus \ldots $ & $2\cdot ({\bf 5},{\bf 2})\oplus 18\cdot ({\bf 3},{\bf 2})\oplus 16\cdot ({\bf 1},{\bf 2})$ \\
\hline 
\end{tabular}
\caption{States composing the subleading $\mathcal{N}=4$ trajectory for different values of $k$.}
\label{table: subleading trajectory}
\end{table}

%%%%%%%%%%%%%%%%%%%%%%%%%%
\subsection{AdS$_{3}\times {\rm S}^{3}\times {\rm K3}$}
%%%%%%%%%%%%%%%%%%%%%%%%%%

One may hope that the situation could become a bit simpler for the case of AdS$_{3}\times {\rm S}^{3}\times {\rm K3}$,
since then the spectrum will contain fewer states. Let us consider the case when K3 can be described as a 
$\mathds{T}^{4}/\mathds{Z}_{2}\,$ orbifold. This $\mathds{Z}_{2}$ orbifold can be easily implemented in the worldsheet description since
it simply acts as a minus sign on each of the four bosonic and fermionic oscillators associated to the $\mathds{T}^4$. 
For each $n$, the surviving states organise themselves into ${\cal N}=4$ multiplets as 
 
\begin{table}[h!]
\centering
\begin{tabular}{c|ccccccccccc}
$s$ & $\delta m=-5$&${-4}$&$-3$&$-2$&$-1$& $0$ & $1$&$2$&$3$&$4$&$5$ \\
\hline
$2n+2$ &  &   &    &    &     & 1  &     &    &    &   & \\
$2n+3/2$ &  &   &    &    & 0   &    & 0   &    &    &   & \\
$2n+1$   &  &   &    & 2  &     & 5  &     & 2  &    &   & \\
$2n+1/2$ &  &   & 0  &    & 10  &    & 10  &    & 0  &   & \\
$2n$   &  & 2 &    & 11 &     & 29 &     & 11 &    & 2 & \\
$2n-1/2$ & 0 &   & 12 &    & 56 &    & 56 &    & 12 &   & 0 \\
$\ldots$ & $\ldots$  & $\ldots$ & $\ldots$   & $\ldots$ & $\ldots$   & $\ldots$ & $\ldots$ & $\ldots$ & $\ldots$  & $\ldots$ \\
\end{tabular}
\caption{Number of the $\mathds{Z}_2$-even $\mathcal{N}=4$ multiplets for $\bar{r}=0$
organized by their $\mathfrak{su}(2)$ quantum numbers. This is to be compared to Table~\ref{table: N = 4 multiplets}.}
\label{table: split 1}
\end{table}

Unfortunately, there is still a fairly large multiplicity (namely $3=5-2$ --- the subtraction of $2$ arises as in the passage from 
Table~\ref{table: N = 4 multiplets} to Table~\ref{table: jp = 0 multiplets}) for the first odd spin `leading' Regge trajectory
states, and again the most natural intepretation is that the leading Regge trajectory has 
just even spin multiplets as before. Similarly, the situation for the subleading Regge trajectory also does not
seem to improve significantly. 

%%%%%%%%%%%%%%%%%%%%%%%%%%%%%%%%%%%%%%%%%%%%%%%%%%%%%%%%%%%%%%%%%%%%
\section{Spectrally flowed sectors and long strings}\label{sec: spectral}
%%%%%%%%%%%%%%%%%%%%%%%%%%%%%%%%%%%%%%%%%%%%%%%%%%%%%%%%%%%%%%%%%%%%
In the previous section we have identified the low-lying states of the leading Regge trajectory that originate from the
unflowed discrete representations. More specifically, these states have spin $s=2n+2$, with $n=0,1,2,\ldots, \frac{1}{4} (k+2)$, 
where the upper bound comes from eq.~(\ref{cons1}), which in turn is a consequence of the unitarity bound (\ref{unitarity bound}).
If we impose the slightly stronger MO-bound (\ref{MO bound}), we find instead 
\begin{equation}\label{spinbound}
s < \frac{k}{2} + 2 - \frac{1}{2k} \ . 
\end{equation}
In either case, we only get finitely many states in this manner. 
In this section we look for the remaining states of the leading Regge trajectory. As we shall see, they arise
from the continuous representations describing long strings. This makes also intuitive sense since the leading Regge trajectory
states correspond to longer and longer strings that get closer to the boundary of AdS$_3$, until they finally merge with the continuum of
long strings.

We start by describing the rest of the full string spectrum that corresponds to the spectrally flowed 
continuous and discrete representations. For each class of representation we then
identify the states of lowest mass for a given spin. We will see that the states from the unflowed discrete representations 
are indeed the lightest states of a given spin for small spin; furthermore, for $s\approx \frac{k}{2}$, the spectrally flowed 
continuous representations will take over. 

The spectrally flowed representations are obtained from the discrete and continuous representations upon applying the 
automorphism of $\sltwof$ defined by 
\begin{equation}\label{spectral flow automorphism}
\begin{alignedat}{4}
\tilde{J}^{3}_{n} &= J^{3}_{n} + \tfrac{k}{2}\omega\delta_{n,0} \\
\tilde{J}^{\pm}_{n} &= J^{\pm}_{n\mp \omega} \\
\tilde{\psi}^{3}_{r} &= \psi^{3}_{r} \\  
\tilde{\psi}^{\pm}_{r} &= \psi^{\pm}_{r\mp \omega} \\ 
\tilde{L}_{n}  &=  L_{n} - \omega J^{3}_{n} - \tfrac{k}{4}\omega^{2}\delta_{n,0} \ .
\end{alignedat}
\end{equation}
Here $\omega$ is an integer, and the same automorphism (with the same value of $\omega$) is applied to both left- and right-movers. We characterise the spectrally
flowed representations by using the same underlying vector space, but letting the $\tilde{J}^a_m$ modes
act on it (rather than the $J^a_m$ modes), and similarly for the fermions. In order for the resulting representation to decompose 
into lowest  weight representations of $\mathfrak{sl}(2,\mathds{R})$ we need, in particular, that 
\begin{equation}\label{lowest weight}
\tilde{J}^-_0 \, |j,m\rangle =  J^-_{\omega}\, |j,m\rangle = 0 \ .
\end{equation}
Thus we take $\omega$ to be a positive integer (or zero). Note that the $\tilde{J}^3_0$ eigenvalue of the states is then
\begin{equation}\label{sCFT}
\tilde{J}^3_0 \, |j,m\rangle = \left( m +  \frac{k}{2}\, \omega \right)\,  |j,m\rangle \ ,
\end{equation}
where $m$ is the actual $J^3_0$ eigenvalue of the state in question. Since $\omega$ is positive, this eigenvalue is always positive 
(at least on the ground states).

Using the explicit form of  $\tilde{L}_{0}$ (c.f. \eqref{spectral flow automorphism}),  the on-shell condition is in the NS-sector and for general $\omega$ 
\begin{equation}\label{onshell}
-\frac{j(j-1)}{k}-\omega\left(m+\frac{k}{4}\omega \right) + N_{\text{tot}} = \frac{1}{2} \ ,
\end{equation}
where $N_{\text{tot}} = N + N' + N''$ is the total excitation number, and we have set $j'=h^{\mathds{T}}=0$. 
A similar condition also applies to the right-movers, and we have the level-matching condition
\begin{equation}\label{levmatch}
N_{\text{tot}}-\bar{N}_{\text{tot}}=\omega s \  
\end{equation}
since (\ref{onshell}) involves the term $\omega m$. Finally, we need to impose the GSO-projection. It is natural to assume --- and this
leads to the correct BPS spectrum of \cite{Argurio:2000tb} --- that the correct GSO projection is the one that takes the same form in all
representations, including the spectrally flowed ones. In terms of the original vector space description we are using here,
this then translates into the condition 
\begin{equation}\label{GSO_flowed}
N_{\text{tot}}+\frac{\omega+1}{2}\in\mathds{N} \ ,
\end{equation}
since we only flow in the $\mathfrak{sl}(2,\mathbb{R})$ factor and hence the fermion number of 
the ground state changes by one for each unit spectral flow, see also \cite{Giribet:2007wp}.\footnote{There is a similar spectral flow
automorphism for $\sutwof$, but since this does not lead to new representations, it is a matter
of convenience whether we include this flow or not. In our context, it is simpler not to flow in this sector.}

%%%%%%%%%%%%%%%%%%%%%%%%%%%%%%%%%%%
\subsection{Spectrally flowed representations --- the continuous case}\label{specflowcont}
%%%%%%%%%%%%%%%%%%%%%%%%%%%%%%%%%%%
According to \cite{Maldacena:2000hw} the spectrum of string theory on AdS$_3$ contains representations
whose ground states transform in continuous representations of $\sltwo$. The states of the continuous representation
${\cal C}_j^\alpha$ are labelled by $|j,m,\alpha\rangle$, where $j=\frac{1}{2} + i p$ with $p$ real, and 
$m$ takes all values of the form $m=\alpha + \mathds{Z}$. 
These representations are neither highest nor lowest weight with respect to $\sltwo$. Their Casimir is given by 
\begin{equation}
C_2\,   |j,m,\alpha\rangle = - j (j-1) \, |j,m,\alpha\rangle \ , \qquad 
- j (j-1) = \frac{1}{4} + p^2 \ . 
\end{equation}
In particular, they can therefore only satisfy the mass-shell condition (\ref{Virasoro constraints}) in the NS-sector with 
$N_{\text{tot}}=\bar{N}_{\text{tot}}=0$. Since this is incompatible with the GSO projection, there are no physical
states in the \emph{unflowed} continuous representations.\footnote{This should better be so, since otherwise the dual CFT would have had
an unbounded $L_0$ spectrum.} However, after spectral flow, these representations give rise to interesting physical states, as
we shall now describe.

Because of \eqref{lowest weight} (applied to $|j,m,\alpha\rangle$) the 
spectrally flowed continuous representations are lowest weight with respect to $\mathfrak{sl}(2,\mathds{R})$ if $\omega>0$. Plugging
$j=\frac{1}{2} + i p$ into the mass-shell condition (\ref{onshell}) and solving for $m$ leads to 
\begin{equation}\label{meq}
m=-\frac{k\omega}{4}+\frac{1}{\omega}\left( N_{\text{tot}} - \frac{1}{2} +\frac{p^2}{k}+\frac{1}{4k} \right) ,
\end{equation}
and similarly for $\bar{m}$. (Remember that $j$, i.e., $p$, and $\omega$ are the same for both left- and right-moving 
representations.) Using (\ref{sCFT}) the spacetime energy of the state is then 
\begin{equation}
E_{\text{cont}} = m+\frac{k\omega}{2}+\bar{m}+\frac{k\omega}{2} 
= \frac{k\omega}{2} + \frac{1}{\omega} \left( N_{\text{tot}} + \bar{N}_{\text{tot}} - 1+\frac{2p^2}{k}+\frac{1}{2k} \right)  \ .
\end{equation}
It is clear that the lowest energy for any given quantum numbers is achieved by putting $p=0$, as also expected classically. 
Furthermore, using level-matching (\ref{levmatch}) to solve for the spin we can rewrite this as 
\begin{equation}
E_{\text{cont}}(s) = s +\frac{k\omega}{2} + \frac{1}{\omega} \left( 2\bar{N}_{\text{tot}} +\frac{1}{2k} -1 \right)  \  .
\end{equation}
Thus the minimum energy for a given $s$ is achieved by putting $\bar{N}_{\text{tot}}=0$ if $\omega$ is odd, or 
$\bar{N}_{\text{tot}}=1/2$ if $\omega$ is even (as required by the GSO projection, 
eq.~(\ref{GSO_flowed})). For any $k > \sqrt{6}/2-1\simeq 0.22$ and any even $\omega \geq 2$, the 
continuous $(\omega-1)$ sector has lower energy. Hence,  in what follows we focus on the case of odd $\omega$. Since 
$\omega s=N_{\text{tot}}$, with $\omega\geq 1$, 
setting $\bar{N}_{\text{tot}} = 0$ is only valid for $s\geq 0$. (Analogously, the lowest energy for $s\leq 0$ is achieved by putting $N_{\text{tot}}=0$, so 
that $\omega s = -\bar{N}_{\text{tot}}$.)
Thus we conclude that the spectrally flowed continuous representations contain states with the dispersion relation
\begin{equation}
E_{\text{cont}} (s) = \vert s\vert +\frac{k\omega}{2}  + \frac{1}{\omega} \left( \frac{1}{2k} -1 \right)   \qquad (\omega \text{ odd})
\end{equation}
for any spin $s$. This energy is a growing function of $\omega\in \mathds{N}$ for any $k>-1 + \sqrt{2} \simeq 0.41$. Since
there is no constraint on the set of spins, the lowest energy for any spin is achieved by putting $\omega =1$, for which
we then find 
\begin{equation}\label{Regge_cont}
E_{\text{cont}} (s) = \vert s\vert +\frac{k}{2}  +  \frac{1}{2k} -1  \  .
\end{equation}

\subsubsection{Massless higher spin fields for $k=1$}\label{sec:5.1.1}

We should note that for $k=1$, (\ref{Regge_cont}) describes massless higher spin states. For this value
($k=\omega=1$), the mass-shell condition (\ref{meq}) (and its right-moving analogue) become simply 
\begin{equation}
m = N_{\rm tot}-\frac{1}{2} \ , \qquad  \bar{m} = \bar{N}_{\rm tot}-\frac{1}{2} \ , 
\end{equation}
while the conformal dimensions of the dual CFT are 
\begin{equation}\label{hhbar}
h = m + \frac{1}{2} = N_{\rm tot} \ , \qquad  \bar{h} = \bar{m} + \frac{1}{2} = \bar{N}_{\rm tot}  \ , 
\end{equation}
and the GSO-projection (\ref{GSO_flowed}) requires now that both $N_{\rm tot}$ and $\bar{N}_{\rm tot}$ should be integers. 
Since there are eight transverse oscillators, there is a stringy growth of massless higher spin fields.  

This phenomenon is the exact analogue of what was found for the bosonic case, where the corresponding 
phenomenon happens for $k_{\rm bos} = 1 +2 = 3$ in \cite{GGH}. (In particular, $k=1$ is also the minimum value
where the massless graviton that arises from the discrete representation with $j=1$ is allowed by the MO-bound (\ref{MO bound}).) 
The theory with $k=1$ describes strings scattering off a single NS5 brane;
while this is formally an ill-defined theory --- the level of the bosonic $\mathfrak{su}(2)$ algebra is negative, $k'_{\rm bos}=1-2=-1$, 
although this conclusion could be avoided if we consider instead of ${\rm AdS}_3 \times {\rm S}^3 \times \mathds{T}^4$
the background ${\rm AdS}_3 \times {\rm S}^3 \times {\rm S}^3 \times {\rm S}^1$, see the comments at the bottom of page 2 
 --- it was argued in \cite{Seiberg:1999xz} that at least some aspects of the theory still make sense. Note that the gap
of the spectrum was predicted in \cite{Seiberg:1999xz} to be 
\begin{equation}
\Delta_0 =  \frac{(k-1)^2}{4k} \ , 
\end{equation}
see eq.~(4.26)  of \cite{Seiberg:1999xz}, and this is reproduced exactly (as in the bosonic case of \cite{GGH}) in our
analysis from the mass-shell condition (\ref{meq}) for $p=0$, $w=1$ and $N_{\rm tot}=0$. 
It was furthermore argued there that the dual CFT should correspond to a symmetric orbifold associated to 
$\mathds{R}^4 \times \mathds{T}^4$. (Here the $\mathds{R}^4$ arises from the ${\rm S}^3$ together with the 
radial direction of ${\rm AdS}_3$ that becomes effectively non-compact in this limit.) 
This is nicely in line with our finding of the massless higher spin fields. In particular, given that the symmetric orbifold
involves an $8$-dimensional free theory, the single particle generators have the same growth behaviour as found
above in (\ref{hhbar}), see \cite{Gaberdiel:2015wpo,Gaberdiel:2015mra}. 

On the other hand, this tensionless limit is different in nature to what one expects from the symmetric orbifold of $\mathds{T}^4$,
see \cite{GGH} for a discussion of this point. In particular, one may expect that these massless higher spin states get lifted
upon switching on R-R flux. It would be interesting to confirm this, using the techniques of \cite{Berkovits:1999im}.

%%%%%%%%%%%%%%%%%%%%%%%%%%%%%%%%%%%
\subsection{Spectrally flowed representations --- the discrete case}\label{subsec:specdisc}
%%%%%%%%%%%%%%%%%%%%%%%%%%%%%%%%%%%
For discrete flowed representations, it follows from the analysis of \cite{Maldacena:2000hw} that $j$ satisfies the MO-bound
\begin{equation}
\frac{1}{2} < j < \frac{k+1}{2}  \ . 
\end{equation}
Writing $m=j+r$, and solving the on-shell condition \eqref{onshell} for $j$, 
we find
\begin{equation}\label{sl2}
j=\frac{1}{2}\Biggl[1 - k\omega + \sqrt{1+ 4k \left( N_{\text{tot}} - r \omega -\frac{\omega +1}{4}\right)}\, \Biggr] \ .
\end{equation}
In addition, we must solve the constraints $r\geq -N$ for $\omega$ odd, and 
$r\geq -N -1/2$ for $\omega$ even. We should note 
that $\omega s=N_{\text{tot}}-\bar{N}_{\text{tot}}$, and $s=r-\bar{r}$, so that 
$N_{\text{tot}}-r\omega = \bar{N}_{\text{tot}}-\bar{r}\omega$. Then $j$ 
is indeed the same for the left- and right-moving sectors.

We first note that $j$ is a decreasing function of $r\,$. The unitarity constraint $j\geq 0$, together with the fact that there is a minimum 
value that $r$ can take as a function of $N$, leads to the existence of a minimum value for the levels 
$N_{\text{tot}}, \bar{N}_{\text{tot}}$ in a given $\omega$ sector, 
which is of the form 
\begin{alignat}{3}
\omega\text{ odd:} & \qquad & N_{\text{tot}}\,,\ \bar{N}_{\text{tot}} & \geq \frac{k \omega ^2+2}{4 \omega +4}  \\
\omega\text{ even:} & \qquad & N_{\text{tot}}\,, \ \bar{N}_{\text{tot}} & \geq \frac{k \omega ^2 + 2}{4 \omega +4}-\frac{2\omega}{4\omega + 4}  \ .
\end{alignat}
Let us then define
\begin{equation}
N_{\text{min}}(k,\omega)
 = \left\{
\begin{array}{rcl}
\displaystyle{
\frac{k \omega ^2+2}{4 \omega +4}}
+b \,,& & \text{if } \omega \text{ is odd} \\ 
\hphantom{space}
\\
\displaystyle{
\frac{k \omega ^2 + 2}{4 \omega +4}-\frac{2\omega}{4\omega + 4}} +b \,,& & \text{if } \omega \text{ is even}
\end{array} 
\right.
\end{equation}
where $0\leq b< 1$ is a bookkeeping device that rounds to the closest upper integer if $\omega$ is odd, or to the closest 
upper half-integer if $\omega$ is even, as required by the GSO-projection of eq.~(\ref{GSO_flowed}). Note that there is no upper bound 
on the levels, on the other hand. 
Furthermore, $N_{\text{min}}(k,\omega)$ is an increasing function of both $k$ and, more importantly, $\omega$. This means 
that the lowest allowed levels appear for $\omega=1$.

As for the spectrally flowed continuous representations (that are analysed in Section~\ref{specflowcont}), the lowest energy states are 
those for which either $N_{\text{tot}}$ or $\bar{N}_{\text{tot}}$ (or both) attain their 
lowest possible values. Let us first fix $\bar{N}_{\text{tot}}=N_{\text{min}}(k,\omega)$. Then by level-matching the spin $s$ is positive or null. 
Furthermore, we fix $\bar{r}=-N_{\text{min}}(k,\omega)  - 1/2$ if $\omega$ is even, 
and $\bar{r}=-N_{\text{min}}(k,\omega)$ if $\omega$ is odd. (Note that this is only possible if $N'=N''=0$, 
namely the internal CFT is not excited; this condition will lead to the analog of the even spin lowest energy states in the unflowed case.) This 
uniquely determines $j$ to be
\begin{equation}
j  =\frac{1}{2} \left(1-k\omega+\sqrt{4 b k (\omega +1)+(k \omega -1)^2} \, \right) 
\end{equation}
for both even and odd $\omega$. We see that when $b=0$ we indeed get $j=0\,$, as expected. The energy is then given by
\begin{align}
E_{\text{disc}} (s) & = s + \frac{k\omega}{2}+\frac{1}{\omega}\left( -\frac{2j(j-1)}{k}+2N_{\text{min}}(k,\omega)-1\right) \nonumber \\
& = s + \sqrt{4 b k (\omega +1)+(k \omega -1)^2}-2 b-\frac{\omega  (k \omega -2)}{2 (\omega +1)} \label{Regge_flow} \, 
\end{align}
for odd $\omega$, with a similar expression for even $\omega$. As for the continuous case, for any even $\omega\geq 2$, the discrete $(\omega-1)$-sector has lower energy. Hence we restrict our attention to the case of odd $\omega$, with energy given by \eqref{Regge_flow}. We find that the lowest energy states of positive helicity $s$ are then given by
\begin{alignat}{6}
\omega\text{ even:} & \qquad & N_{\text{tot}} & = N_{\text{min}}(k,\omega)  + \omega s & \qquad & 
m = j - N_{\text{min}}(k,\omega)  - \frac{1}{2}+ s
 \nonumber \\
& \qquad & \bar{N}_{\text{tot}} & = N_{\text{min}}(k,\omega)   & \qquad & \bar{m} = j - N_{\text{min}}(k,\omega)  - \frac{1}{2}
 \nonumber \\
\omega\text{ odd:} & \qquad & N_{\text{tot}} & = N_{\text{min}}(k,\omega)  + \omega s & \qquad & 
m = j - N_{\text{min}}(k,\omega)  + s
 \nonumber \\
& \qquad & \bar{N}_{\text{tot}} & = N_{\text{min}}(k,\omega)   & \qquad & \bar{m} = j - N_{\text{min}}(k,\omega)  \, .
\end{alignat}
We should note that the left-moving states do not saturate the value of $m$ for the given value of $N_{\text{tot}}$, and thus they may have 
multiplicities greater than one. The right-moving states, on the other hand, saturate them, and hence will be unique. Finally, in the
above we have assumed $s>0$; the corresponding lightest states with negative helicity are obtained upon exchanging the roles of 
left- and right-movers.

Even though it is perhaps not evident, the leading energy \eqref{Regge_flow} is an increasing function of $\omega$, 
a fact which we have confirmed numerically. Therefore, the lowest energy states for any given spin come from the 
$\omega =1$ sector. Their energy is
\begin{equation}\label{disclowest}
E_{\text{disc}} (s) =  \vert s\vert  + \sqrt{8 b k+(k-1)^2}-2 b-\frac{k}{4}+\frac{1}{2}  \ .
\end{equation}
We emphasize that the parameter $b$ introduced above is uniquely fixed by $k$ and does not depend on the spin $s$. 
As a result, this dispersion relation is linear in the spin $s$. The same is also true for the states from the spectrally flowed 
continuous representations, see eq.~(\ref{Regge_cont}). This behaviour ties in nicely with the observation
of \cite{AndrejStepanchuk:2015wsq}, see also \cite{Banerjee:2015qeq}, about the behaviour of 
classical strings for large spin $s$. In particular, it is argued in  \cite{AndrejStepanchuk:2015wsq},
see eq.~(6.0.8), that the $\log s$ term correction term to the linear dispersion relation vanishes
for pure NS-NS flux.\footnote{These claims are somewhat in tension with the analysis of 
\cite{David:2014qta} where a $(\log s)^2$ correction term was found for the case of pure NS-NS 
flux. Our findings seem to support the conclusion of \cite{AndrejStepanchuk:2015wsq,Banerjee:2015qeq}.
We thank Arkady Tseytlin for drawing our attention to the work of \cite{AndrejStepanchuk:2015wsq}.}

%%%%%%%%%%%%%%%%%%%%%%%%%%%%%%%%%%%
\subsection{Comparison of the different sectors}
%%%%%%%%%%%%%%%%%%%%%%%%%%%%%%%%%%%
We can now compare the different dispersion relations coming from the different sectors. Recall from the analysis of 
Section~\ref{sec:LR}, see eq.~(\ref{bredge}), that the dispersion relation for the leading Regge trajectory states from 
the unflowed discrete representations is 
\begin{equation}\label{Regge}
E_{\text{Regge}} (s) =1+\sqrt{1+2k (s-2)} \ , 
\end{equation}
where we have set $s=2n+2$ in eq.~(\ref{bredge}) --- this corresponds to the top component of the corresponding 
${\cal N}=4$ multiplet --- and expressed $n$ in terms of $s$. These states are only available for spins up 
$s < \frac{k}{2} + 2 - \frac{1}{2k}$, see eq.~(\ref{spinbound}). (Note that, for $k\geq 2$, the right-hand-side of this 
inequality is not an integer and 
hence cannot be attained.\footnote{For $k=1$, it gives $s=\pm 2\,$.}) It is easy to see that (in this range of spins)
$E_{\text{Regge}} (s)$ from eq.~(\ref{Regge}) is smaller than both $E_{\text{cont}} (s)$ from eq.~(\ref{Regge_cont}) or
$E_{\text{disc}} (s)$ from eq.~(\ref{disclowest}); in fact, precisely at the (unphysical) value $s = \frac{k}{2} + 2 - \frac{1}{2k}$ we 
have 
\begin{equation}
E_{\text{Regge}}  \bigl(s = \tfrac{k}{2} + 2 - \tfrac{1}{2k} \bigr) = 1 + k = 
E_{\text{cont}} \bigl(s = \tfrac{k}{2} + 2 - \tfrac{1}{2k}\bigr) \ . 
\end{equation}
Thus the states from the unflowed discrete representations describe the leading Regge states for spins 
$s < \frac{k}{2} + 2 - \frac{1}{2k}$.

For larger spins, on the other hand, the relevant states must come from the spectrally flowed representations. As we have seen 
in sections~\ref{specflowcont} and~\ref{subsec:specdisc}, for both the spectrally flowed continuous
and discrete representations, the lowest energy states always appear for $\omega=1$, and in either case, they
give rise to states of arbitrarily high spin. We can compare the relevant dispersion relations, and it is fairly
straightforward to see from eqs.~(\ref{Regge_cont}) and (\ref{disclowest}) that 
\begin{equation}
E_{\text{cont}} (s) < E_{\text{disc}} (s) 
\end{equation}
for all spins. Thus it follows that the remaining states of the leading Regge trajectory are part of the
spectrally flowed continuous representations. In order to get a sense of the qualitative picture, we have
plotted the relevant states in Figure~\ref{Regges} for one representative value of $k$ ($k=20$).

The picture that emerges is thus that the lowest energy states arise in the unflowed discrete sector for as high spin as 
allowed by the MO-bound. Once the MO-bound is reached, the continuous $\omega =1$ representations take over; 
this makes intuitive sense since the leading Regge trajectory states come from highly spinning strings that get
longer and longer as the spin is increased. As they hit the boundary of ${\rm AdS}_3$, they merge into the continuum
of long strings \cite{Maldacena:2000hw}, and thus the leading Regge trajectory states of higher spin will arise from 
that part of the spectrum, i.e., from the spectrally flowed continuous representations.

%%%%%%%%%%%%%%%%%%%%%%%%%%%%%%%%%%%%%%%%%%%%%%%%%%%%%%%%%%%%%%%%%%%%
\section{Conclusions}\label{sec: Conclusions}
%%%%%%%%%%%%%%%%%%%%%%%%%%%%%%%%%%%%%%%%%%%%%%%%%%%%%%%%%%%%%%%%%%%%
In this paper we have studied string theory on the background ${\rm AdS}_3 \times {\rm S}^3 \times \mathds{T}^4$ 
with pure NS-NS flux, using the WZW model worldsheet description with a view to exhibiting 
the emergence of a higher spin symmetry in the tensionless (small level) limit. As we have shown in 
Section~3, this part of the moduli space does not contain a conventional tensionless point where small string
excitations become massless and give rise to a Vasiliev higher spin theory. However, for $k=1$, a stringy massless
higher spin spectrum emerges from the spectrally flowed continuous representations (corresponding to long strings). 
These higher spin fields  are of a different nature than those arising in the symmetric orbifold of $\mathds{T}^4$ \cite{GGH}, 
but they realise nicely some of the predictions of \cite{Seiberg:1999xz}. 

For generic values of $k$ we could also identify quite convincingly the states that make up the leading Regge trajectory, 
and we saw that they comprise the spectrum of a Vasiliev higher spin theory with ${\cal N}=4$ superconformal symmetry.  
It would be very interesting to try to repeat the above analysis using the worldsheet description of 
\cite{Berkovits:1999im} that allows for the description of the theory with pure R-R flux (where one would expect 
the actual higher spin symmetry to emerge, see the arguments of the Introduction). Among other things one
should expect that the massless higher spin fields that arise from the long string spectrum at $k=1$
will acquire a mass since the long string spectrum is believed to be a specific feature of the pure NS-NS background. On the 
other hand, the leading Regge trajectory states  should become massless as one flows to the theory
with pure R-R flux. It would be very interesting to confirm these expecations. 
It would also be  very interesting to analyse to which extent the leading Regge trajectory forms a 
closed subsector of string theory in the tensionless limit. 
\smallskip

\acknowledgments
It is a pleasure to thank Marco Baggio, Shouvik Datta, Jan de Boer, Lorenz Eberhardt, Diego Hofman, Chris Hull, Wei Li, 
Charlotte Sleight, Massimo Taronna, and in particular Rajesh
Gopakumar, for useful discussions. MRG thanks the Galileo Galilei Institute for Theoretical Physics (GGI) for the 
hospitality and INFN for partial support during the completion of this work, within the program ``New Developments in 
AdS3/CFT2 Holography''. JIJ thanks the University of Amsterdam String Theory Group, and the Nordic Institute for Theoretical Physics (NORDITA) within the program ``Black Holes and Emergent Spacetime'', for their kind hospitality during the course of this work. This research was also (partly) supported by the NCCR 
SwissMAP, funded by the Swiss National Science Foundation. 

%%%%%%%%%%%%%%%%%%%%%%%%%%%%%%%%%%%%%%%%%%%%%%%%%%%%%%%%%%%%%%%%%%%%
%%%%%%%%%%%%%%%%%%%%%%%%%%%%%%%%%%%%%%%%%%%%%%%%%%%%%%%%%%%%%%%%%%%%

\appendix
%%%%%%%%%%%%%%%%%%%%%%%%%%%%%%%%%%%%%%%%%%%%%%%%%%%%%%%%%%%%%%%%%%%%
\section{Supersymmetric current algebras: conventions and useful formulae}\label{app: SUSY Kac Moody}
%%%%%%%%%%%%%%%%%%%%%%%%%%%%%%%%%%%%%%%%%%%%%%%%%%%%%%%%%%%%%%%%%%%%

The ${\cal N}=1$ superconformal WZW model is generated by a bosonic Kac-Moody algebra $\mathfrak{g}$, 
coupled to fermions in the adjoint representation 
\begin{align}
J^{a}(z)J^{b}(w) 
\sim{}&
 \frac{if^{ab}_{\hphantom{ab}c}J^{c}(w)}{z-w} + \frac{k\eta^{ab}}{(z-w)^{2}}
\\
J^{a}(z)\psi^{b}(w)
\sim{}&
 \frac{if^{ab}_{\hphantom{ab}c}\psi^{c}(w)}{z-w}
 \\
 \psi^{a}(z)\psi^{b}(w)  
 \sim{}&
 \frac{k\eta^{ab}}{z-w}\ .
\end{align}
In terms of modes,
\begin{align}
\left[ J^a_m, J^b_n \right] 
&=
 if^{ab}_{\hphantom{ab}c}J^c_{m+n}+km\eta^{ab}\delta_{m,-n}
  \\
\left[ J^a_m, \psi^b_r \right] 
&=
 if^{ab}_{\hphantom{ab}c}\psi^c_{m+r}
  \\
\left\lbrace \psi^a_r, \psi^b_s \right\rbrace 
&=
  k\,\eta^{ab}\delta_{r,-s}\ .
\end{align}
The structure constants satisfy $f^{ab}_{\hphantom{ab}c} = -f^{ba}_{\hphantom{ba}c}$ by definition; moreover,
$f^{abc}$ can be chosen to be completely anti-symmetric (for a semi-simple Lie algebra). In addition, the Jacobi identity reads
\begin{equation}\label{Jacobi identity}
f^{ab}_{\hphantom{ab}d}f^{dc}_{\hphantom{dc}e}+f^{ca}_{\hphantom{ca}d}f^{db}_{\hphantom{db}e}
+f^{bc}_{\hphantom{bc}d}f^{da}_{\hphantom{da}e}
=
0\ ,
\end{equation}
and the dual Coxeter number $h^\vee$ may be defined through
\begin{equation}\label{Coxeter}
f^{d}_{\hphantom{d}bc}f^{abc} = 2\, h^\vee\, \eta^{ad}\,.
% \qquad
%  \bigl(\Rightarrow \quad f^{d}_{\hphantom{d}bc}f^{ac}_{\hphantom{ac}d} = 2\check{h} \delta^{a}_{\hphantom{a}b}\,,\quad f^{d}_{\hphantom{d}bc}f^{ab}_{\hphantom{ab}d} = -2 
%\check{h}\delta^{a}_{\hphantom{a}c}\bigr).
\end{equation}
We can decouple the fermions from the bosons by defining the shifted currents $\cJ^{a}$ as
\begin{equation}
\cJ^{a}
=
J^{a}+\frac{i}{2k}f^{a}_{\hphantom{a}bc}\bigl(\psi^{b}\psi^{c}\bigr)\ ,
\end{equation}
where the round brackets denote normal ordering. Because of the anti-symmetry of the structure constants, this is trivial in the NS-sector,
but there is a subtle contribution in the R-sector since we have --- we follow the conventions of \cite{Fuchs:1992nq}, see in particular
eq.~(3.1.43), 
\begin{align}
(\psi^{a}\psi^{b})_{p}  ={}&
 \frac{1}{2}\, [\psi^{a}_{0},\psi^{b}_{p} ]
 +\sum_{m \leq -1}\psi^{a}_{m}\psi^{b}_{p-m}-\sum_{m \geq 1}\psi^{b}_{p-m}\psi^{a}_{m} \ .  \label{Rordering}
\end{align}
With this definition, the OPEs become
\begin{align}
\cJ^{a}(z)\cJ^{b}(w) 
\sim{}&
 \frac{if^{ab}_{\hphantom{ab}c}\cJ^{c}(w)}{z-w} + \frac{\sk\, \eta^{ab}}{(z-w)^{2}}
\\
\cJ^{a}(z)\psi^{b}(w)
\sim{}&
0 \ ,
\end{align}
where 
%\begin{equation}
$\sk \equiv k- h^\vee$, 
%\end{equation}
and $h^\vee$ is the dual Coxeter number defined in \eqref{Coxeter}. Equivalently, in terms of modes we find
\begin{align}
\left[ \cJ^{a}_{m}, \cJ^{b}_{n} \right] 
&=
 if^{ab}_{\hphantom{ab}c} \,\cJ^c_{m+n}+\sk\, m\eta^{ab}\delta_{m,-n} \\
\left[ \cJ^{a}_{m}, \psi^{b}_r \right] 
&= 
0\ .
\end{align}
Hence, the algebra is isomorphic to the direct (commuting) sum of a bosonic affine algebra at shifted level $\kappa$, 
and $\text{dim}(\mathfrak{g})$ free fermions.

Using the above shifted currents we obtain the stress tensor and a dimension-3/2 supercurrent via the Sugawara construction,
\begin{align}\label{Sugawara currents 1}
T ={}&
 \frac{1}{2k}\eta_{ab}\left[ (\cJ^{a}\cJ^{b})  - ( \psi^{a}\partial \psi^{b}) \right]
 \\
 G ={}&
 \frac{1}{k}\left[\eta_{ab}\,  \cJ^{a}\psi^{b} - \frac{if_{abc}}{6k}\bigl(\psi^{a}\psi^{b}\psi^{c}\bigr)\right] \ , 
 \label{Sugawara currents 2}
\end{align}
where the round brackets denote normal-ordering, 
and the triple product is defined recursively, i.e., $(\psi^{a}\psi^{b}\psi^{c}) \equiv (\psi^a (\psi^b \psi^c))$.\footnote{Because 
of the total anti-symmetry of the structure constants, normal ordering is again trivial in the NS-sector, but there is a contribution 
coming from the commutator term in (\ref{Rordering}).}
% through \eqref{multiple normal ordering}. The Sugawara currents \eqref{Sugawara currents 1}-\eqref{Sugawara currents 2} satisfy the $\cN=1$ superconformal algebra
These fields satisfy the ${\cal N}=1$ superconformal algebra 
\begin{align}
\label{N equals 1 superconformal algebra 1}
T(z)T(w) 
\sim {}&
\frac{c/2}{(z-w)^{4}} + \frac{2T(w)}{(z-w)^{2}} + \frac{\partial T(w)}{z-w}
\\
T(z)G(w)
 \sim{}&
 \frac{3}{2}\frac{G(w)}{(z-w)^{2}} + \frac{\partial G(w)}{z-w}
\\
G(z)G(w) 
\sim{}&
\frac{2c/3}{(z-w)^{3}} +\frac{2T(w)}{z-w}
\label{N equals 1 superconformal algebra 3}
\end{align}
with central charge 
\begin{equation}\label{total central charge}
c = \text{dim}(\mathfrak{g})\left(\frac{k-h^\vee}{k}+ \frac{1}{2}\right)= \text{dim}(\mathfrak{g})\left(\frac{\sk}{\sk +h^\vee}+ \frac{1}{2}\right).
\end{equation}
In terms of modes, in the NS sector we have
\begin{align}
\bigl[L_{m},L_{n}\bigr]
={}&
(m-n)L_{m+n} + \frac{c}{12}m\left(m^{2}-1\right)\delta_{m,-n}
\\
\bigl[L_{n},G_{r}\bigr] 
={}&
\left(\frac{n}{2}-r\right)G_{n+r}
\\
\bigl\{G_{r},G_{s}\bigr\}
={}&
2L_{r+s} + \frac{c}{3}\left(r^{2}-\frac{1}{4}\right)\delta_{r,-s}\,.
\end{align}
Due to the non-trivial R-sector normal ordering term, see eq.~(\ref{Rordering}), for the above definition
of the normal ordered modes we obtain in the R-sector the algebra
%(c.f. \eqref{normal ord modes General}), 
\begin{align}\label{R N=1 algebra v1}
\bigl[L_{m},L_{n}\bigr]
%={}&
%(m-n)L_{m+n} +m \left( \frac{c}{12}\left(m^{2}-1\right) +\frac{\text{dim}(\mathfrak{g})}{8}\right)\delta_{m,-n}
%\nonumber\\
={}&
(m-n)L_{m+n} +m\frac{\text{dim}(\mathfrak{g})}{8} \left(m^{2} - \frac{2 h^\vee}{3k}(m^{2}-1)\right)\delta_{m,-n}
\\
\bigl[ L_n,G_r \bigr] ={}&
 \left( \frac{n}{2}-r \right) G_{r+n} 
\\
\bigl\{ G_{r}, G_{s} \bigr\} ={}&
 2L_{r+s} + \frac{c}{3}s^{2}\delta_{r,-s}+\frac{h^\vee\text{ dim}(\mathfrak{g})}{12k}\delta_{r,-s}
 \label{R N=1 algebra v1 2}
\end{align}
instead. Note in particular that the Virasoro commutator $\left[L_m,L_n\right]$ does not have the standard form. If so desired, this can be 
rectified by shifting the zero mode of the stress tensor as
\begin{equation}
L_{n} \to L_{n}^{R} = L_{n}  +  \frac{\text{dim}(\mathfrak{g})}{16}\delta_{n,0}\ ,
\end{equation}
so that the superconformal algebra then reads 
\begin{align}
\Bigl[L^{R}_{m},L^{R}_{n}\Bigr]
={}&
(m-n)L^{R}_{m+n} + \frac{c}{12}m\left(m^{2}-1\right)\delta_{m,-n}
\\
\Bigl[ L^{R}_{n},G_{r} \Bigr] ={}&
 \left( \frac{n}{2}-r \right) G_{n+r} 
\\
\Bigl\{ G_{r}, G_{s} \Bigr\} ={}&
 2L^R_{r+s} + \frac{c}{3}\left( r^{2} - \frac{1}{4}\right)\delta_{r,-s} \ ,
 \end{align}
which is exactly as in the NS sector. The price one pays for this redefinition is that the fermionic Ramond vacuum 
$\vert 0\rangle_{\text{R}}$ (which is annihilated by all the positive modes of the fermions) is no longer annihilated by $L_{0}$, but 
rather satisfies
\begin{equation}
L_{0}^{R}\vert 0\rangle_{\text{R}} = \frac{\text{dim}(\mathfrak{g})}{16} \, \vert 0\rangle_{\text{R}}
\end{equation}
instead.

 Finally, it is interesting to note that if we simultaneously consider supersymmetric $\sltwo$ and 
$\mathfrak{su}(2)$ algebras (which have $h^\vee$ equal and opposite in sign), as appropriate to AdS$_{3}\times {\rm S}^{3}$, 
we find that the $h^\vee$ terms in \eqref{R N=1 algebra v1} -- \eqref{R N=1 algebra v1 2} drop out from the algebra of the total currents.

%%%%%%%%%%%%%%%%%%%%%%%%%%%%%%%%%%%%%%%%%%%%%%%%%%%%%%%%%%%%%%%%%%%%
\section{Low momenta subtleties}\label{app:physstatessubtleties}
%%%%%%%%%%%%%%%%%%%%%%%%%%%%%%%%%%%%%%%%%%%%%%%%%%%%%%%%%%%%%%%%%%%%

The only subtlety concerning the counting of physical states given by eqs.~(\ref{characters}) and  (\ref{characters 2})
arises for $j=1$ and $j'=0$. Then the mass-shell condition requires that the physical states appear at excitation number
$N=\frac{1}{2}$, and in particular, the state that is excited by $\psi^{-}_{-1/2}$ has $j=0$. For $j=0$ the general character
formula for $\mathfrak{sl}(2,\mathds{R})$ representations (\ref{sl2char}) breaks down since the $L_{-1} = J^+_0$ descendant
of the state with $j=0$ is null. As a consequence, we have the identity 
\begin{equation}
\frac{y^0}{1-y} = 1 + \frac{y}{1-y} = \chi_{j=0} + \chi_{j=1} \ ,
\end{equation}
i.e., the character on the left-hand-side is actually not an irreducible character, but rather splits up into the contributions
of two different irreducible $\mathfrak{sl}(2,\mathds{R})$ representations (namely the ones with $j=0$ and $j=1$). This
phenomenon also has a microscopic origin: for $j=1$ and $j'=0$ there are three $\mathfrak{sl}(2,\mathds{R})$ 
descendants that define physical states, namely 
\begin{align}
\bigl|\tfrac{1}{2};2,2\bigr\rangle 
={}&
  \psi^{-}_{-1/2}|1,3\rangle -6\psi^{3}_{-1/2}|1,2\rangle +6\psi^{+}_{-1/2}|1,1\rangle
\\
\bigl|\tfrac{1}{2};1,1\bigr\rangle 
={}&
 \psi^{-}_{-1/2}|1,2\rangle -2\psi^{3}_{-1/2}|1,1\rangle 
\\
\bigl|\tfrac{1}{2};0,0\bigr\rangle 
={}&
 \psi^{-}_{-1/2}|1,1\rangle \ , 
\end{align}
and one easily confirms that all three of them are physical. (And, in fact, 
$\bigl|\tfrac{1}{2};1,1\bigr\rangle = J^+_0 \, \bigl|\tfrac{1}{2};0,0\bigr\rangle$, reflecting
the null-vector structure mentioned before.) 

Since one of these states ($\bigl|\tfrac{1}{2};0,0\bigr\rangle$) is the vacuum state of the space-time CFT,
these states describe the chiral states of the space-time CFT. (Recall that the above discussion is a chiral
discussion; the vacuum state for the right-movers, say, appears then together with the above states.) In
particular, the $j=h=2$ state is the
Virasoro field, and at $j=h=1$ we get in addition to the state $\bigl|\tfrac{1}{2};1,1\bigr\rangle$ 
six $j=h=1$ states from the excitations associated to the ${\rm S}^3 \times \mathds{T}^4$ directions. 
Altogether, they give rise to an $\mathfrak{su}(2)$ current algebra (coming from the ${\rm S}^3$ excitations), 
as well as four $h=1$ bosons --- these are the familiar bosons of the $\mathds{T}^4$. (Similary, in 
the R-sector we get four $h=\frac{1}{2}$ fields and four $h=\frac{3}{2}$ fields --- they describe the four 
free fermions of the $\mathds{T}^4$, as well as the four supercharges of the ${\cal N}=4$ 
superconformal algebra.)

%%%%%%%%%%%%%%%%%%%%%%%%%%%%%%%%%%%%%%%%%%%%%%%%%%%%%%%%%%%%%%%%%%%%
\section{The structure of small $\mathcal{N}=4$ multiplets}\label{app: multiplets}
%%%%%%%%%%%%%%%%%%%%%%%%%%%%%%%%%%%%%%%%%%%%%%%%%%%%%%%%%%%%%%%%%%%%

The (small) ${\cal N}=4$ superconformal algebra is generated by a Virasoro algebra with modes $L_n$, an affine $\sutwo$ algebra
with modes $T^a_n$ (where $a=\pm,3$), as well as four supercharges $Q^{i \pm}$ where $i=1,2$. (The supercharges 
transform as two doublets with respect to the $\sutwo$ algebra).  The commutation relations are of the form 
\begin{align}
\left[L_{m},L_{n}\right] 
={}&
 (m-n)L_{m+n} + \frac{c}{12}m(m^{2}-1)\delta_{m+n,0}
\nonumber\\
\left[L_m,T^{a}_{n}\right] 
={}&
 -nT^{a}_{m+n} 
\nonumber\\
\left[L_m,Q^{i\, \pm}_{n}\right] 
={}&
 \left(\tfrac{m}{2}-n\right)Q^{i\, \pm}_{m+n}
\nonumber\\
\left[T^{3}_{m},T^{\pm}_{n}\right] 
={}&
 \pm  T^{\pm}_{m+n}
\nonumber\\
\left[T^{3}_{m},T^{3}_{n}\right] 
={}&
 \frac{c}{12}m\delta_{m+n,0}
\nonumber\\
\left[T^{+}_{m},T_{n}^{-}\right] 
={}&
2\, T^{3}_{m+n} + \tfrac{c}{6}m\delta_{m+n,0}
\\
\left[T^{3}_{m},Q^{i \pm}_{n}\right] 
={}&
 \pm \frac{1}{2}Q^{i \pm}_{m+n}
 \ ,\qquad \qquad 
 \left[T^{\pm}_{m},Q^{i\pm}_{n}\right]
 =
 0
\nonumber \\
\left[T^{\pm}_{m},Q^{1\mp}_{n}\right] 
={}&
 -Q^{1\pm}_{m+n}\ ,
 \qquad \qquad
 \left[T^{\pm}_{m},Q^{2\mp}_{n}\right] 
=
 Q^{2\pm}_{m+n}
 \nonumber\\
 \left\{Q_{m}^{i \pm},Q_{n}^{i\pm}\right\} 
 ={}&
  \left\{Q_{m}^{i \pm},Q_{n}^{i\mp}\right\} 
=  0 
 \nonumber\\
 \left\{Q^{1\pm}_{m},Q^{2\mp}_{n}\right\}
 ={}&
 2L_{m+n} \pm 2(m-n)T^{3}_{m+n} + \tfrac{c}{3}\left(m^{2}-\tfrac{1}{4}\right)\, \delta_{m+n,0}
 \nonumber\\
  \left\{Q^{1\pm}_{m},Q^{2\pm}_{n}\right\}
 ={}&
-2(m-n)T^{\pm}_{m+n} 
\nonumber
\end{align}
where the central charge is $c=6k$. We denote the corresponding right-moving generators as 
$\bar{L}_n$, $\bar{Q}^{i\pm}$, and $\bar{T}^a_n$. 

With these preparations we can now describe the structure of the supermultiplets. We shall first
concentrate on the chiral (say left-moving) algebra and describe the representations of the 
(small) $\mathcal{N}=4$ superconformal algebra, keeping track of the $\mathfrak{su}(2)$ 
quantum numbers. Following the notation in  \cite{deBoer:1998ip}, we will label  $\mathfrak{su}(2)$ 
representations by their dimension $m = 2j' + 1$. A generic multiplet has then the form, see 
Table~\ref{table: chiral long multiplet}.

\begin{table}[h!]
\centering
\begin{tabular}{|c|c|c|}
\hline
state		& $\mathfrak{su}(2)$  & $h$ \\
\hline
$\vert m\rangle$		 		 				& {\bf m} 								& $h$  \\
$Q\vert m\rangle$ 				& $2\cdot ({\bf m+1}) \oplus 2\cdot ({\bf m-1})$ 		& $h+1/2$\\
$QQ\vert m\rangle$ & 4 $\cdot {\bf m} \oplus ({\bf m+2}) \oplus ({\bf m-2})$		& $h+1$  \\
$QQQ\vert m\rangle$ 				& $2\cdot ({\bf m+1}) \oplus 2\cdot ({\bf m-1})$ 		& $h+3/2$ \\
$QQQQ\vert m\rangle$ 				& {\bf m} 		& $h+2$ \\
\hline
\end{tabular}
\caption{Chiral long $\mathcal{N}=4$ multiplet.}
\label{table: chiral long multiplet}
\end{table}

For small values of ${\bf m}$, namely ${\bf m}=1$ and ${\bf m}=2$, there are various shortenings; more specifically, we find 
for ${\bf m}=1$ and ${\bf m}=2$ the shorter multiplets, see Table~\ref{table: chiral long multiplet vacuum}.

\begin{table}[h!]
\centering
\begin{tabular}{|c|c|c||c|c|c|}
\hline
state					  			  		 & $\mathfrak{su}(2)$ 					 & $h$ 
& state & $\mathfrak{su}(2)$ 					 & $h$  \\
\hline
$\vert 1\rangle$		 		 				& {\bf 1} 								& $h$ & 
$\vert 2\rangle$		 		 				& {\bf 2} 								& $h$  \\
$Q\vert 1\rangle$ 				& $2 \cdot {\bf 2}$  & $h+1/2$ 
& $Q\vert 2\rangle$ 				& $2 \cdot {\bf 3} \oplus 2 \cdot {\bf 1}$  & $h+1/2$ \\
$QQ\vert 1\rangle$ & ${\bf 3} \oplus 3 \cdot {\bf 1}$ 		& $h+1$ 
& $QQ\vert 2\rangle$ & ${\bf 4} \oplus 4 \cdot {\bf 2}$ 		& $h+1$  \\
$QQQ\vert 1\rangle$ 				& $2 \cdot {\bf 2}$ 	& $h+3/2$ 
& $QQQ\vert 2\rangle$ 				& $2 \cdot {\bf 3} \oplus 2 \cdot {\bf 1}$ 	& $h+3/2$ \\
$QQQQ\vert 1\rangle$ 				& {\bf 1} 		& $h+2$ 
& $QQQQ\vert 2\rangle$ 				& {\bf 1} 		& $h+2$  \\
\hline
\end{tabular}
\caption{Chiral long $\mathcal{N}=4$ multiplet for ${\bf m}=1$ and ${\bf m}=2$.}
\label{table: chiral long multiplet vacuum}
\end{table}

If we denote the highest weight state as $\vert h;j'\rangle$, the $\frac{1}{4}$ BPS bound for the above algebra is obtained by demanding that 
$Q^{i+}_{-1/2}\, \vert h;j'\rangle=0$ for one choice of $i\in\{1,2\}$. Using that 
\begin{equation}
\left\{Q^{2-}_{1/2},Q^{1+}_{-1/2}\right\} = \left\{Q^{1-}_{1/2},Q^{2+}_{-1/2}\right\} = 2\left(L_0-T^{3}_{0}\right) \ , 
\end{equation}
we see that 
every $\frac{1}{4}$ BPS state is automatically $\frac{1}{2}$ BPS, i.e., if $Q^{i+}_{-1/2}\, \vert h;j'\rangle=0$ for one choice of $i$, 
it is actually zero for both $i=1,2$. Furthermore, the BPS bound is explicitly 
\begin{equation}
\text{BPS bound:}\qquad \left(L_{0} - T^{3}_{0}\right)\, \vert h;j'\rangle =0 \qquad \Longrightarrow \qquad h = j' \ .
\end{equation}
The resulting short multiplet is described in table \ref{table: chiral BPS multiplet}. As usual, for small values of $j'$ (or ${\bf m}$), there
are further shortenings, in particular, for ${\bf m}=1$ the whole multiplet consists just of the vacuum itself $h=j'=0$, while
for ${\bf m}=2$ the whole multiplet truncates to ${\bf 2}\,  (h=\frac{1}{2}) \oplus 2\cdot {\bf 1}\, (h=1)$.

\begin{table}[h!]
\centering
\begin{tabular}{|c|c|c|}
\hline
state					  			  		 & $\mathfrak{su}(2)$ 					& $h$ \\
\hline
$\vert m\rangle$		 		 				& {\bf m} 											& $j'$ \\
$Q\vert m\rangle$ 				&  $2 \cdot ({\bf m-1})$ 		& $j'+1/2$ \\
$QQ\vert m\rangle$ & ${\bf m -2}$ 		& $j'+1$  \\
\hline
\end{tabular}
\caption{Chiral BPS multiplet. (Recall that $m=2j'+1$.)}
\label{table: chiral BPS multiplet}
\end{table}

The corresponding multiplets of the full $(4,4)$ theory is then obtained by tensoring these chiral multiplets together. 
For example, if both left- and right-moving multiplets are long (corresponding to ${\bf m}$ and $\bar{\bf m}$), 
the total number of states is $256 \times m \cdot \bar{m}$.

%%%%%%%%%%%%%%%%%%%%%%%%%%%%%%%%%%%%%%%%%%%%%%%%%%%%%%%%%
%%%BIBLIOGRAPHY
%%%%%%%%%%%%%%%%%%%%%%%%%%%%%%%%%%%%%%%%%%%%%%%%%%%%%%%%%

\providecommand{\href}[2]{#2}\begingroup\raggedright\endgroup

\end{document}